\documentclass[fleqn,usenatbib]{mnras}
\usepackage{graphicx}
\usepackage{amsmath,amssymb,amsfonts}
\usepackage[cal=cm]{mathalfa}
\usepackage{pgfplots} 
\usepackage{smartdiagram}
\pgfplotsset{compat = newest} 
\DeclareRobustCommand{\VAN}[3]{#2}
\let\VANthebibliography\thebibliography
\def\thebibliography{\DeclareRobustCommand{\VAN}[3]{##3}\VANthebibliography}

\def \la{Lyman alpha}

\def\vx{\hbox{\bf x}}

\newcommand{\be}{\begin{equation}}
\newcommand{\ee}{\end{equation}}
\newcommand{\daa}{\Delta\alpha/\alpha}

\title[VPFIT theory]{Precision in high resolution absorption line modelling, analytic Voigt derivatives, and optimisation methods.}

\author[Webb, Carswell, Lee]{
John K. Webb$^1$\thanks{jkw.phys@gmail.com},
Robert F. Carswell$^2$\thanks{rfc@ast.cam.ac.uk},
Chung-Chi Lee$^1$\thanks{lee.chungchi16@gmail.com}.
\\ \\
$^{1}$Clare Hall, University of Cambridge, Herschel Rd, Cambridge CB3 9AL.\\
$^{2}$Institute of Astronomy, University of Cambridge, Madingley Road, Cambridge, CB3 0HA, UK.\\
}

\date{Accepted \phantom{mmmmmmm}. Received \phantom{mmmmmmm}; in original form \phantom{mmmmmmm}}
\pubyear{2021}

\begin{document}
\label{firstpage}
\pagerange{\pageref{firstpage}--\pageref{lastpage}}
\maketitle

\begin{abstract}
This paper describes the optimisation theory on which {\sc vpfit}, a non-linear least-squares program for modelling absorption spectra, is based. Particular attention is paid to precision. Voigt function derivatives have previously been calculated using numerical finite difference approximations. We show how these can instead be computed analytically using Taylor series expansions and look-up tables. We introduce a new optimisation method for an efficient descent path to the best-fit, combining the principles used in both the Gauss-Newton and Levenberg-Marquardt algorithms. A simple practical fix for ill-conditioning is described, a common problem when modelling quasar absorption systems. We also summarise how unbiased modelling depends on using an appropriate information criterion to guard against over- or under-fitting.

The methods and the new implementations introduced in this paper are aimed at optimal usage of future data from facilities such as ESPRESSO/VLT and HIRES/ELT, particularly for the most demanding applications such as searches for spacetime variations in fundamental constants and attempts to detect cosmological redshift drift.
\end{abstract}

\begin{keywords}
quasars: absorption lines, cosmology: observations, methods: data analysis
\end{keywords}

\section{Introduction}

{\sc vpfit} \citep{ascl:VPFIT2014} is an optimisation code designed primarily for the analysis of high resolution quasar spectra. However it has been used in a number of other applications, including the interstellar medium, absorption lines in stellar photospheres, and emission line fitting. {\sc vpfit} has been cited in more than 300 papers\footnote{ADS Abstracts citations since 1995; actual usage is higher.}. A comprehensive {\sc vpfit} user guide is given in \citep{web:VPFIT}. The mathematical details for a fledgling version of {\sc vpfit} were provided in \cite{Webb1987} but have not been reported in the peer-reviewed literature.

This paper first summarises the theoretical basis for {\sc vpfit} and then describes new enhancements and inclusions that improve accuracy, stability, and provide quantitative information about systematic uncertainties. Modifications of particular note are (i) new analytic calculations of Voigt function derivatives (previously finite difference derivatives were used) and analytic derivatives for all other (non-Voigt) parameters, (ii) addition of distortion parameters inside the non-linear least squares processes, (iii) enhanced resolution and interpolation in the Voigt function look-up tables, and (iv) the hybridisation of Gauss-Newton and Levenberg-Marquardt to form a unified new optimisation method.

The new modifications are in part motivated by the recent application of Artificial Intelligence methods \citep{Bainbridge2017, gvpfit2017, Lee2020AI} and Information Criterion techniques \citep{Webb2021} to spectroscopy. The advances reported here facilitate optimal analyses of future high signal to noise and high calibration precision data achievable with new and forthcoming spectroscopic facilities\footnote{Notably, the Echelle SPectrograph for Rocky Exoplanets and Stable Spectroscopic Observations (ESPRESSO) on the European Southern Observatory's Very Large Telescope (VLT) \citep{espresso2021} and the High Resolution Echelle Spectograph (HIRES) on the forthcoming Extremely Large Telescope (ELT) e.g. \citet{Marconi2016, ELT2018}.}, especially future challenges such as redshift drift and searches for varying fundamental constants.

High precision in computing the Voigt function and profile is paramount. Sections \ref{sec:2algorithms} and \ref{sec:lookuptable}) discuss precision and also introduce a modified optimisation method, merging the two different approaches used in the Gauss-Newton and Levenberg-Marquardt techniques.

Accurate derivatives of the Voigt profile are also essential for optimal, robust, $\chi^2$ descent and unbiased parameter estimation. Previous computations (in {\sc vpfit} and, as far as we know, in other codes) have either been based on finite difference approximations, or have made use of analytic approximations (which lack accuracy or are computationally demanding or both). Finite difference derivatives ({\it fdd}) can work well although have two important disadvantages: (i) {\it fdd} intervals need to be chosen according to the characteristics of the data being modelled, and (ii) the chosen intervals may in fact not be appropriate for all absorption components within a complex comprising many absorption lines (since line parameters and blending vary substantially). 

In this paper we introduce a new approach for calculating Voigt derivatives that entirely eliminates such difficulties. We use Taylor series expansions of the derivatives of the Voigt function, with look-up tables. This method is analogous to that of \cite{Harris1948} applied to the Voigt function itself, except now the idea is applied directly to {\it analytic derivatives} of the Voigt function. Section \ref{sec:derivs} describes this and makes use of the derivative convolution theorem to allow for instrumental resolution. Both of the problems mentioned above are then avoided, no user decisions are needed, and the new method provides significant precision and some speed improvements compared to finite difference derivatives or analytic approximations. Section \ref{sec:discussion} summarises the advances described in this work and two appendices discuss several practical aspects of the calculations.

\section{Non-linear least-squares minimisation} \label{sec:2algorithms}

A comprehensive description of non-linear optimisation methods is given in the excellent book by \citet{GMW81}. The application of such methods to the detection and measurement of stars in crowded fields was described in \cite{Irwin1985}. Both of the previous citations were strongly influential in the application of non-linear optimisation to spectroscopy described in this paper.

Non-linear least-squares methods fall into two broad classes: the simpler Gradient methods, using only first order derivatives and Newton-type methods which use both first and second derivatives of an objective function. The latter are more powerful since they generally converge faster and are more robust. The rate at which a Newton method converges depends on the form of the function being minimised. The nearer the function is to quadratic, the faster the convergence. If it is exactly quadratic, convergence can be achieved in a single iteration. When residuals are Gaussian, non-linear least squares techniques are equivalent to Maximum Likelihood methods and provide optimal parameter estimates \citep{Charnes1976}. An additional important advantage of Newton-type methods is that reliable parameter error estimates are available at virtually no extra computing effort.

Let the model intensity be $I(\mathbfit{x})$, where the vector $\mathbfit{x}$ is the set of all free model parameters and let the $i^{th}$ normalised residual between model and data $d_i$ (having uncertainties $\sigma_i$) be
\be
f(\mathbfit{x})_i = \frac{I(\mathbfit{x})_i-d_i}{\sigma_{i}} \, .
\label{eq:f(x)i}
\ee
where the subscript $i$ is the index in the spectral array at which the observed-frame wavelength is $\lambda_i$. $I(\mathbfit{x})$ is the model intensity after convolution with the instrumental resolution $\mathcal{G}(\lambda)$ in wavelength space, i.e.
\be
I(\mathbfit{x})_i = (I_{\nu} \ast \mathcal{G})_i = \frac{\int_{-\infty}^\infty I_{\nu} \mathcal{G}(\lambda)_i d\lambda}{ \int_{-\infty}^\infty \mathcal{G}(\lambda)_i d \lambda} \, ,
\label{eq:Ix}
\ee
where $I_{\nu}$ is calculated from Eq.~\eqref{eq:absorption}, expressed as a function of wavelength as defined by Eq.~\eqref{eq:u}, and $\mathbcal{G}(\lambda)_i$ is the instrumental profile for pixel-$i$. {\sc vpfit} provides the option of $\mathbcal{G}$ being either a Gaussian instrumental profile or a user-defined numerical instrumental profile\footnote{As required by Hubble Space Telescope spectroscopic data for example, and as is likely to be required for future high-precision spectroscopy with instruments such as ESPRESSO/VLT and HIRES/ELT}. 

To fit $I(\mathbfit{x})$ to the set of $n$ observed data points $d_i$, we want to minimise
\be
F(\mathbfit{x}) = \frac{1}{2} \sum_{i=1}^n f(\mathbfit{x})_i^2 = \frac{1}{2} \mathbfit{f(x)}^T \mathbfit{f(x)} \, ,
\label{eq:chisq}
\ee
where $T$ denotes transpose and the factor of 1/2 has been included to eliminate an extra factor of 2 in subsequent derivatives of this equation. 

To set up the minimisation procedure, we approximate $F(\mathbfit{x})$ using a quadratic model, i.e. a Taylor series expansion to second order,
\begin{align}
F(\mathbfit{x} + \mathbfit{p}) & \approx F(\mathbfit{x}) + \mathbfit{p}^T f'(\mathbfit{x}) + \frac{1}{2} \, \mathbfit{p}^T f''(\mathbfit{x}) \, \mathbfit{p} \nonumber \\
& = F(\mathbfit{x}) + \mathbfit{p}^T g(\mathbfit{x}) + \frac{1}{2} \, \mathbfit{p}^T \mathbfss{G}(\mathbfit{x}) \, \mathbfit{p} \, ,
\label{eq:taylor}
\end{align}
where $\mathbfit{p}$ is the predicted parameter unit vector update that minimises $F(\mathbfit{x} + \mathbfit{p})$, prime denotes derivative, $g(\mathbfit{x})$ is the gradient vector, and $\mathbfss{G}(\mathbfit{x})$ is the Hessian matrix. For the $q^{\mathrm{th}}$ model parameter, the corresponding component of the gradient vector is
\be
g(\mathbfit{x})_q = \frac{\partial F(\mathbfit{x})}{\partial x_q} = \sum_{i=1}^n \frac{\partial I(\mathbfit{x})_i}{\partial x_q} f(\mathbfit{x})_i \, ,
\label{eq:gradient}
\ee
or in vector/matrix form,
\be
g(\mathbfit{x}) = \mathbfss{J}(\mathbfit{x})^T \mathbfit{f(x)}
\label{eq:gradvec} \, ,
\ee
where $\mathbfss{J}(\mathbfit{x})$ is the $n \times m$ Jacobian matrix of $\mathbfit{f(x)}$ ($n$ is the number of data points, $m$ is the number of free parameters) whose $i^{th}$ row is
\be
\nabla f(\mathbfit{x})_i = (\partial f_i/\partial x_1, \partial f_i/\partial x_2, \dots, \partial f_i/\partial x_m) \, .
\label{eq:Jacobian}
\ee

For any two model parameters $x_q$ and $x_r$, the component of the Hessian matrix is
\begin{align}
& G(\mathbfit{x})_{qr} = \frac{\partial^2F(\mathbfit{x})}{\partial x_q \partial x_r} \nonumber \\
& \quad = \left[ \sum\limits_{i=1}^n \frac{\partial^2 I(\mathbfit{x})_i}{\partial x_q \partial x_r} f(\mathbfit{x})_i \right] + \left[ \sum\limits_{i=1}^n \frac{\partial I(\mathbfit{x})_i}{\partial x_q} \frac{\partial I(\mathbfit{x})_i}{\partial x_r} \frac{1}{\sigma_i^2}  \right] \, ,
\label{eq:hessian}
\end{align}
or in vector/matrix form,
\be
\mathbfss{G}(\mathbfit{x}) = \mathbfss{Q}(\mathbfit{x}) + \mathbfss{J}(\mathbfit{x})^T \mathbfss{J}(\mathbfit{x}) \, .
\label{eq:hessianmatrix}
\ee

Consider the first term in Eq.~\eqref{eq:hessian}. If the model $I(\mathbfit{x})$ is a reasonable representation of the data, each $f(\mathbfit{x})_i$ may be considered as an independent random variable such that $\langle f(\mathbfit{x})_i \rangle \rightarrow 0$ as $n \rightarrow \infty$. For this reason, the second term in Eq.~\eqref{eq:hessian} dominates, so we drop the first term in square brackets above, and the Hessian matrix can be approximated using only first order derivatives,
\be
\mathbfss{G}(\mathbfit{x}) \approx \mathbfss{J}(\mathbfit{x})^T \mathbfss{J}(\mathbfit{x}) \, . \label{eq:hessianapprox}
\ee

An advantage of approximating the Hessian using only first-order derivatives, apart from simplicity, is that it renders $\mathbfss{G}(\mathbfit{x})$ positive-definite, thereby guaranteeing a descent direction when solving the matrix equations for optimal parameter updates. To minimise Eq.~\eqref{eq:taylor} by the choice of some suitable $\mathbfit{p}$, it is convenient to formulate a quadratic function in terms of $\mathbfit{p}$, the step to the minimum, rather than the predicted minimum itself. Then at each iteration (i.e. for some particular set of model parameters $\mathbfit{x}$ at the current iteration), the optimal parameter updates are found by minimising Eq.~\eqref{eq:taylor} with respect to $\mathbfit{p}$. Doing so gives
\be
g(\mathbfit{x}) = -\mathbfss{G}(\mathbfit{x}) \, \mathbfit{p}_{\mathrm{min}} \, ,
\label{eq:gGp}
\ee
An algorithm in which the search direction is obtained using equations of the form \eqref{eq:gGp} is called a “Newton type method”. The equation used in {\sc vpfit} is not \eqref{eq:gGp}, but instead is a modified form of it, for the reasons discussed shortly in Sections \ref{sec:hybrid} to \ref{sec:compare}. For simplicity, from here on we drop the ``$(\mathbfit{x})$'' notation.

\subsection{Gauss-Newton and Levenberg-Marquardt methods} \label{sec:hybrid}

The two best known non-linear minimisation methods are Gauss-Newton (GN) and Levenberg-Marquardt (LM). We have experimented using both individually but also have most recently implemented a hybrid procedure formed from both.

The GN method attempts to improve efficiency by tweaking the search direction at each iteration. After solving Eq.~\eqref{eq:gGp} for $\mathbfit{p}_{\mathrm{min}}$, an extra univariate minimisation is carried out, to identify that value of $\alpha$ which minimises $F(\mathbfit{x} + \alpha \mathbfit{p}_{\mathrm{min}})$. The Gauss-Newton parameter updates are then
\be
\mathbfit{p}_{\mathrm{GN}} = \alpha \mathbfit{p}_{\mathrm{min}} \, . \label{eq:GN}
\ee

The LM method takes a different approach, modifying the Hessian matrix,
\be
g = -\mathbfss{G}_{LM} \mathbfit{p}_{\mathrm{min}} = -\left( \mathbfss{G} + \eta \mathbcal{I} \right) \mathbfit{p}_{\mathrm{min}} \, , \label{eq:LM}
\ee
where $\mathbcal{I}$ is an identity matrix and $\eta$ is a non-negative scalar, adjusted iteratively to find the largest reduction in $F(\mathbfit{x} + \mathbfit{p}_{\mathrm{min}})$. 
Newton's method is recovered when $\eta=0$ and when $\eta\mathbcal{I} \gg \mathbfss{G}$, the search direction becomes parallel to that of the gradient descent method (but step lengths are altered by a factor of $1/\eta$). Improvements to the standard LM process have been proposed, for problems in which the number of parameters is large, e.g. \citet{Transtrum2012}, but we have not explored those particular methods.

\subsection{Switching between GN and LM}

Empirically, at some points during minimisation, GN can produce a larger step towards convergence, whilst at other points, LM does so. In {\sc vpfit} version 12.2 and earlier, GN and LM are used in conjunction with each other, based on the Hessian described by Eq.~\eqref{eq:finalhessian}; GN and LM parameter updates are computed at every {\sc vpfit} iteration and the descent direction was taken to be that which produces the largest drop in chi squared, $\Delta F$. In practice, the extra time lost in computing both GN and LM descents was more than offset by a greater convergence efficiency. To assist the following discussion, we call this switching procedure the ``GN-LM'' method.

\subsection{Hybrid Optimisation (HO), ill-conditioning, and Modified Cholesky Factorisation (MCF)}\label{sec:ill-conditioning}

As will be seen shortly, when the GN method is modified to account for ill-conditioning, it is very similar to LM. This, and the previous discussion, suggests what seems to be an obvious point - instead of the GN-LM approach (computing both GN and LM solutions and then chosing the best at each iteration), one could simply unify GN and LM to form a single minimisation technique in the following way: within the same iteration, first use Eq.~\eqref{eq:LM} to solve for $\eta$, then use Eq.~\eqref{eq:GN} to find $\alpha$ (or {\it vice versa}). This unification of the GN and LM methods can result in a faster descent than either GN or LM individually and is implemented in {\sc vpfit} version 12.3. Since individually, GN and LM each have only one tuning parameter each, the hybrid method has two. To assist the following discussion, we will refer to the new hybrid method as the ``Hybrid Optimisation (HO) method''. However, prior to solving for the descent direction using this new HO approach, it is necessary to first deal with dynamic range and ill-conditioning problems.

In principle, the parameter updates $\mathbfit{p}_{\mathrm{min}}$ could be obtained by solving Eq.~\eqref{eq:gGp} using Cholesky decomposition. However, practical issues arise that require modifications of Eq.~\eqref{eq:gGp}:
\begin{enumerate}
    \item The quadratic model, Eq.~\eqref{eq:taylor}, is unlikely to be perfect, particularly when far from the best-fit solution;
    \item $\mathbfss{G}$ often has a huge dynamical range. The reason is easy to understand; redshift parameters are generally far more tightly constrained than column density or $b$-parameters. There are similar considerations for other parameters, such that rounding errors can become important when using Cholesky decomposition to solve for $\mathbfit{p}_{\mathrm{min}}$. Empirically, the dynamic range can be $\sim 10^{14}$ or even greater, creating precision difficulties. If left unchecked, the consequence of this can be to render solutions of Eq.~\eqref{eq:gGp} unstable;
    \item Ill-conditioning is often present, again causing stability problems when applying Cholesky decomposition to Eq.~\eqref{eq:gGp}.
\end{enumerate} 
Therefore, the following modifications are carried out. As described in Section \ref{sec:2algorithms}, the Hessian matrix and gradient vector are formed from derivatives with respect to the free parameters (Eq.~\eqref{eq:gradient} and \eqref{eq:hessian}). The solution implemented in {\sc vpfit} (prior to applying the HO method) involves two modifications of the Hessian matrix:\\ 
{\bf (1)} Each row/column in the Hessian is normalised such that all diagonal terms are unity,
\be
\mathbfss{G}_n = \mathbfss{D}^{-1/2} \mathbfss{G} \mathbfss{D}^{-1/2} \, ,
\label{eq:unitdiag}
\ee
where the subscript ``$n$'' indicates normalised, $\mathbfss{D}_{ij} = \sqrt{\mathbfss{G}_{ii}} \delta_{ij}$. This reduces the dynamic range, lessening the impact of possible rounding error problems, and is also useful for ill-conditioning.\\
{\bf (2)} We assume ill-conditioning occurs (it frequently does, particularly when line blending inevitably results in some parameters being poorly determined). The Hessian is rendered positive-definite by adding a constant to its diagonal terms. The theoretical basis for this is discussed in e.g. \cite{GMW81}. The solution implemented in {\sc vpfit}, the second modification of the Hessian, is
\be
\mathbfss{G}_{VPFIT} \rightarrow \mathbfss{G}_n + \eta_n \mathbcal{I} \, ,
\label{eq:finalhessian}
\ee
where $\eta_n$ is a tunable non-negative scalar and $\mathbcal{I}$ is the identity matrix.

\begin{figure}
\centering
\input{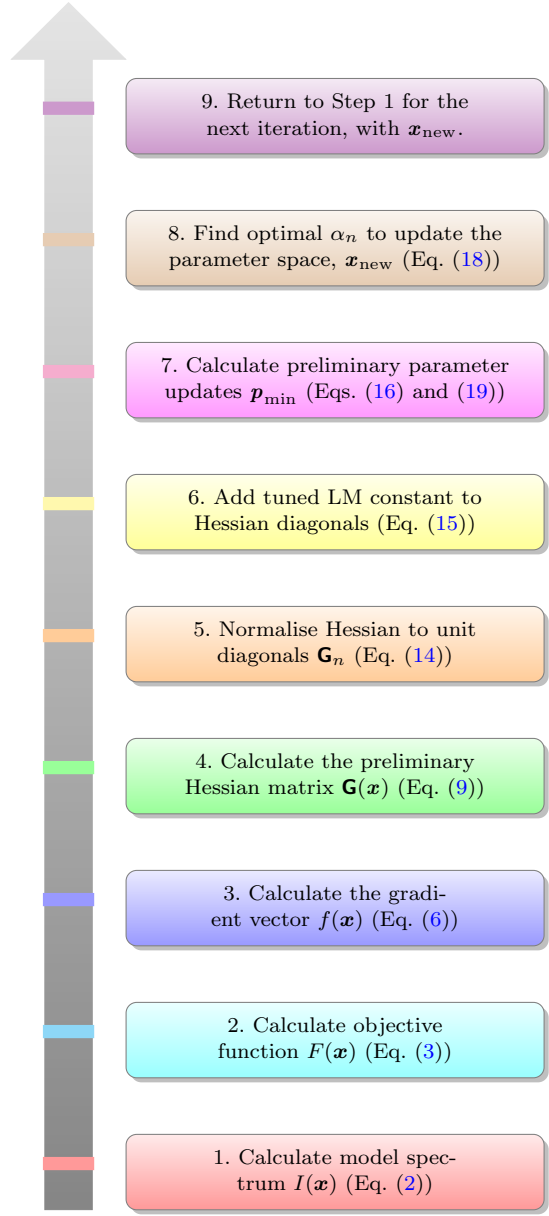}
\caption{Bottom-up flow chart illustrating sequential logic for one {\sc vpfit} iteration.}
\label{fig:vpchart}
\end{figure}

As Eq.~\eqref{eq:LM} shows, MCF and LM are in fact similar. Both add constants to the Hessian diagonals. To maximise the iterative reduction in $F(\mathbfit{x} + \mathbfit{p})$, MCF (as implemented here) adds the tunable constant to the normalised Hessian whilst LM does the same but using the unnormalised Hessian. MCF (again, as implemented here) doubles the diagonal terms. The advantage (of both approaches) is that the positive-definite Hessian guarantees a descent direction so the process is always stable. The penalty is that in modelling situations that are not inherently ill-conditioned, MCF reduces efficiency such that reaching the minimum requires a slightly larger number of steps. The final matrix equations, i.e. the modification to Eq.~\eqref{eq:gGp}, therefore become
\be
g^n = -(1+\eta_n) \mathbfss{G}_n \, \mathbfit{p}^n_{\mathrm{min}} \, , \label{eq:gGp_n}
\ee
where
\be
g^n = \mathbfss{D}^{-1/2} g \quad \mathrm{and} \quad \mathbfit{p}^n_{\mathrm{min}} = \mathbfss{D}^{1/2} \mathbfit{p}_{\mathrm{min}} \,.
\ee
Having now modified the Hessian in the two ways just discussed and then solved for $\mathbfit{p}_{\mathrm{min}}$ using MCF, the final step in our hybrid procedure is univariate minimisation of $F$ by optimising $\alpha_n$ to find the parameter updates,
\be
\mathbfit{x}_\mathrm{new} = \mathbfit{x} + \alpha_n \mathbfss{D}^{-1/2} \mathbfit{p}^n_{\mathrm{min}} \,,
\label{eq:GN_n}
\ee
i.e. the new hybrid procedure described above tunes two new parameters $\eta_n$ and $\alpha_n$ to minimise $F$\footnote{In practice, $\eta_n$ is tuned quite coarsely, stepping in powers of $10$. This has the advantage of speed and empirically it works well. In principle however, finer tuning could result in faster descent per iteration but would require more computing time within each iteration. We have not yet explored the trade-off between these two things.}. Because of the way in which the new HO method has been set-up, {\it from the same set of starting parameters within a single iteration}, it must always descend at least as rapidly than LM or GN and in general will win. However, HO will follow a different descent path than either LM or GN and therefore a more efficient overall descent is not guaranteed. The logic flow for one {\sc vpfit} iteration is illustrated schematically in Figure \ref{fig:vpchart}.

\subsection{Comparing GN-LM and HO methods} \label{sec:compare}

We can thus compare the search directions obtained using the new hybrid method to those obtained using the regular GN and LM methods. To do so let us first transfer Eq. \eqref{eq:gGp_n} back by multiplying by $\mathbfss{D}^{1/2}$, which gives
\be
g = -\left( \mathbfss{G} + \eta_n \mathbfss{D} \right) \, \mathbfit{p}_{\mathrm{min}} \, , \label{eq:gGp_n2}
\ee
showing that when $\eta_n = 0$, the HO method reduces to the GN method and as $\eta_n \rightarrow \infty$, the search direction (solved for using Eq.~\eqref{eq:gGp_n2}) becomes $\mathbfit{p}_{\mathrm{min}} \propto \mathbfss{D}^{-1} g$, such that the impact of off-diagonal terms becomes negligible, i.e. we assume that each parameter is parabolic along its respective axis in parameter space.

\begin{figure}
\centering
\includegraphics[width=0.98\linewidth]{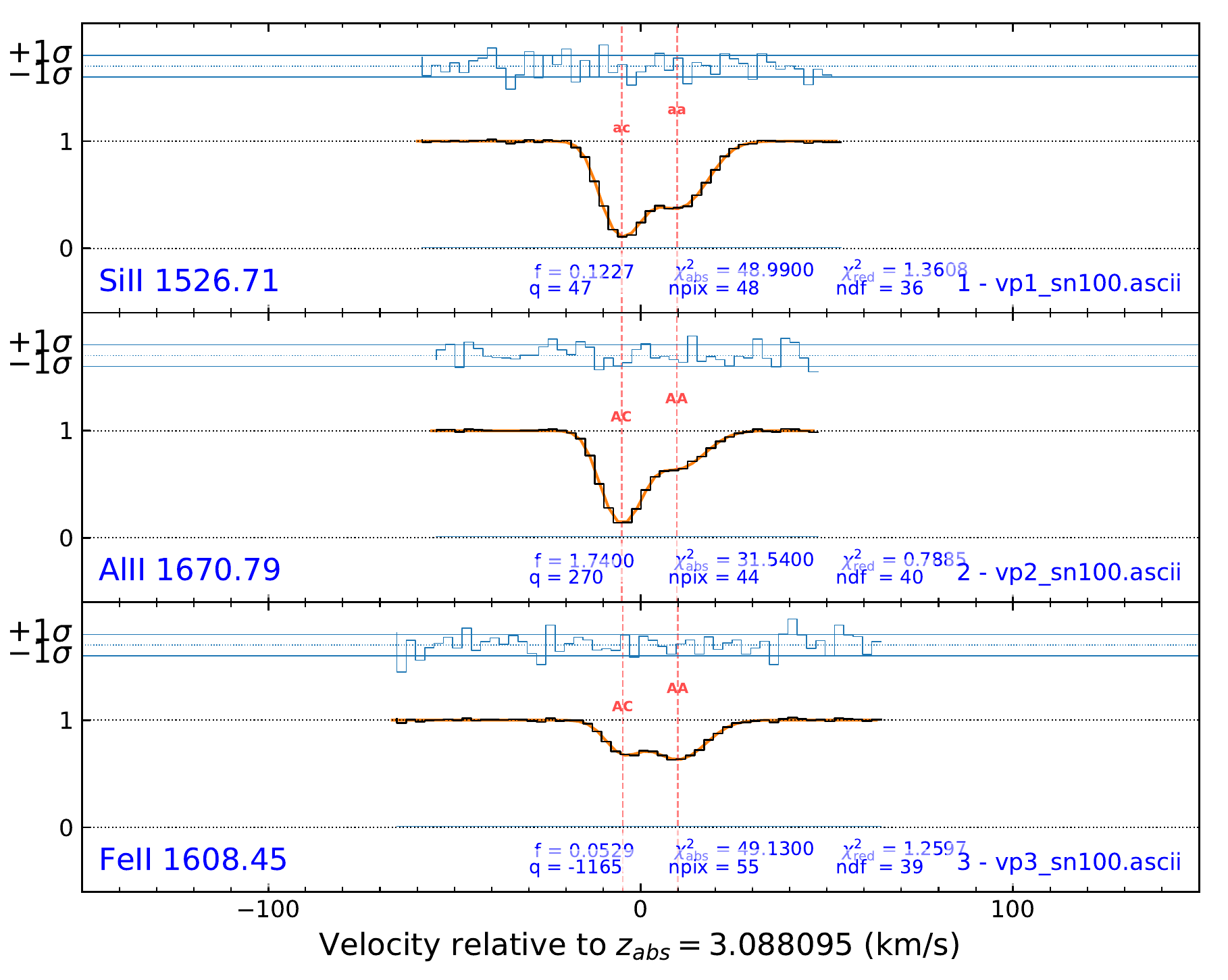}
\caption{Synthetic spectrum of a two-component absorption system, with three atomic species. The signal to noise per pixel is 100 and the spectral resolution is 6 km/s FWHM. The broadening mechanism is turbulent i.e. all transitions at the same redshift have the same $b$-parameter. The data were fitted twice, each time with a different set of starting guesses.}
\label{fig:Model_t}
\end{figure}

The relative performances of the hybrid and GN-LM methods will, of course, vary according to the data being modelled, the starting parameter guesses, and other observational details. Figure \ref{fig:Model_t} illustrates a two-component synthetic absorption system, comprising three atomic species. The signal to noise is 100 per pixel. Absorption line parameters taken from a known absorption system towards the bright $z_{em}=3.12$ quasar Q0420-388. The three spectral segments were fitted simultaneously using {\sc vpfit}. The fine structure constant was included as a free parameter. The data were fitted twice, each time using a different set of starting guesses, which were perturbed far from the true input parameter values (initial normalised chi squared values for the starting models were $\sim 700$ and $\sim 900$).

\begin{figure}
\centering
\includegraphics[width=0.98\linewidth]{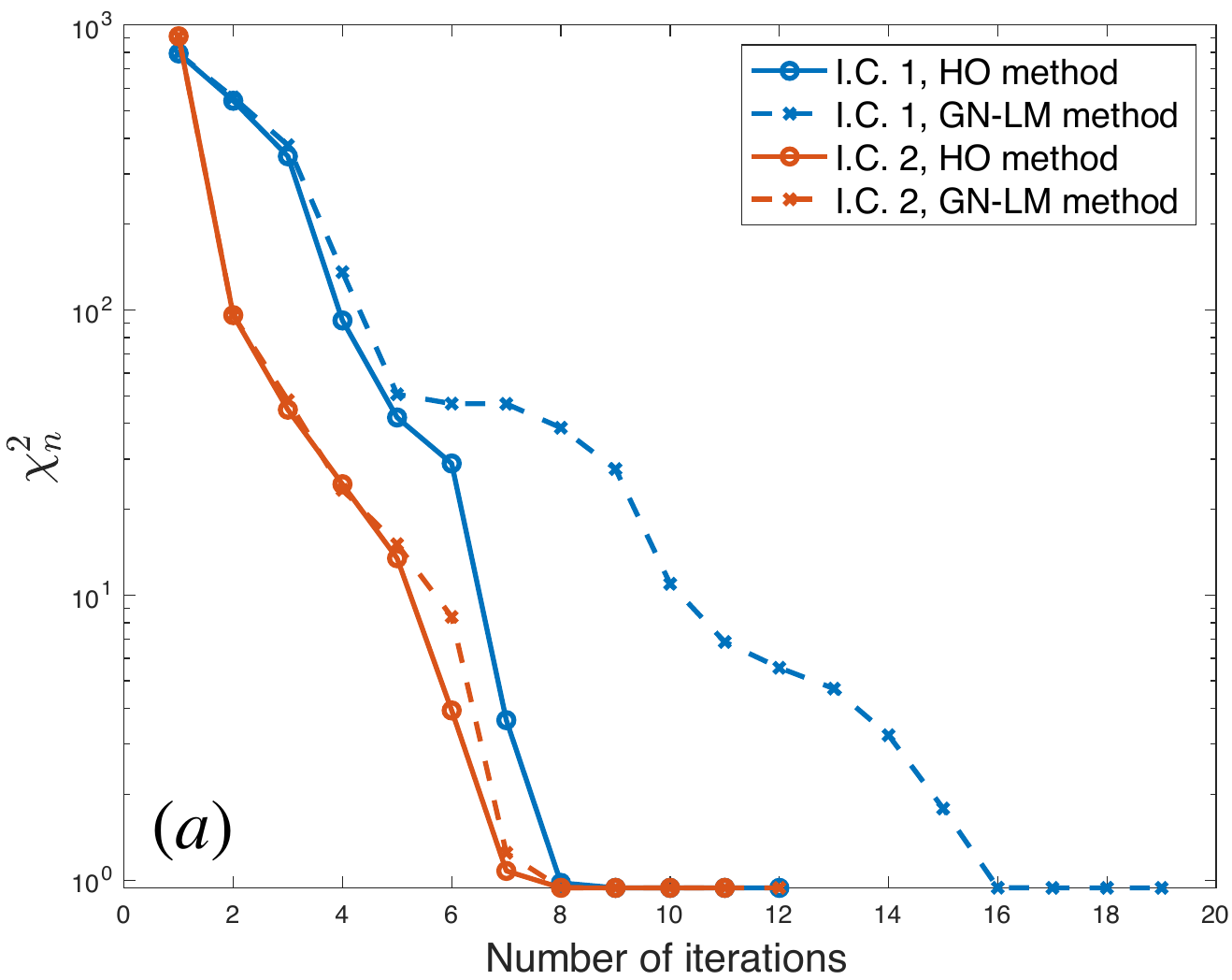} \\
\includegraphics[width=0.98\linewidth]{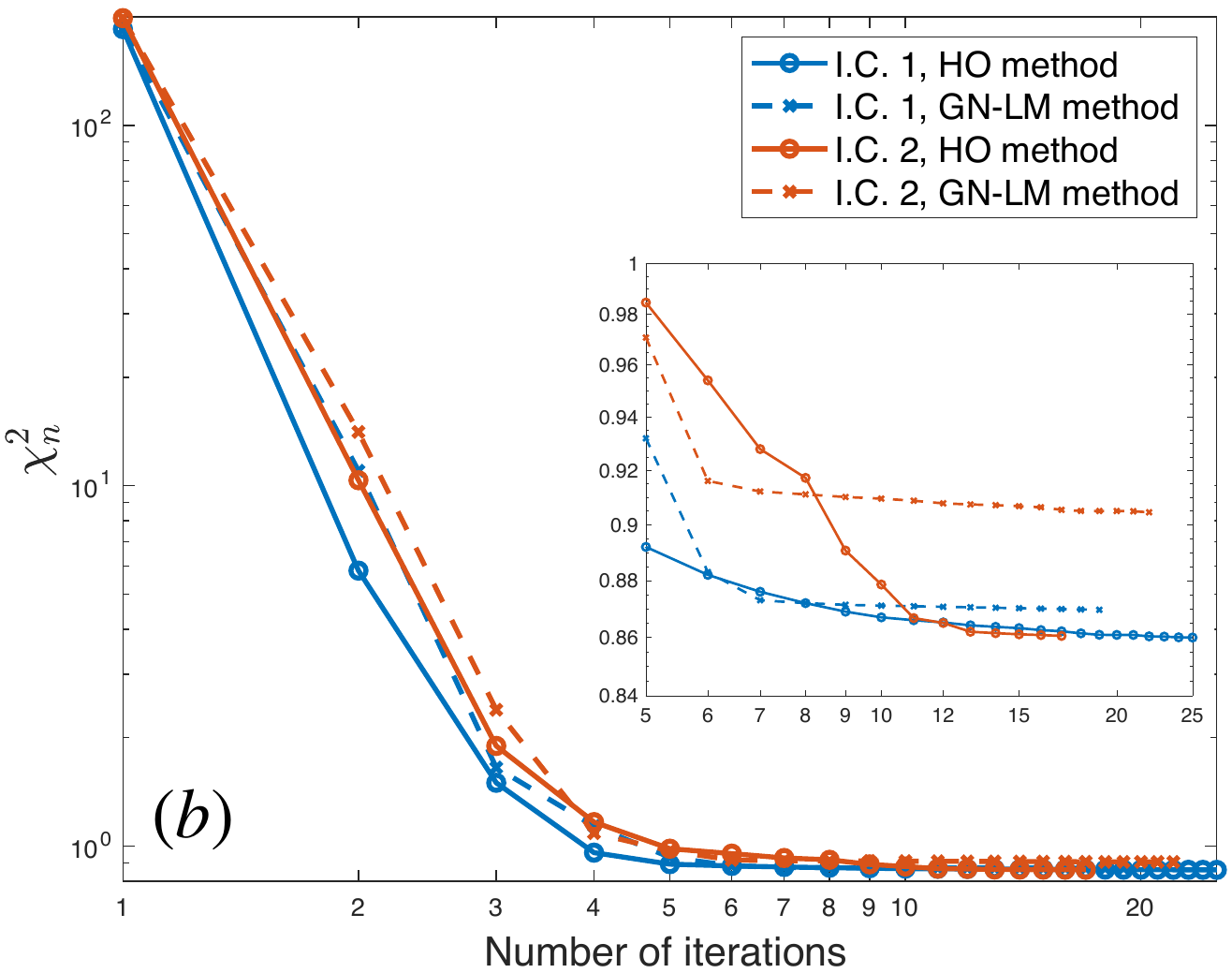}
\caption{The evolution of the normalised $\chi^2_n$ per iteration for trial fits, each with a different set of starting guesses. The solid lines illustrate descent for the HO optimisation method and the dashed lines are for the GN-LM method. Panel (a) shows the results for fitting a synthetic spectrum (described in Section \ref{sec:compare}). Panel (b) shows the results for fitting a real spectrum.}
\label{fig:HO-GN-LM}
\end{figure}

Figure \ref{fig:HO-GN-LM} shows the evolution of the normalised value of $\chi^2_n$ at each iteration ($\chi^2_n = 2F(\mathbfit{x})/ndf$, where $ndf$ is the number of degrees of freedom in the fit). We examine the relative performances of the hybrid and GN/LM methods by running both methods on a synthetic spectrum (panel (a)) and also on real data (panel (b)). Different transitions were used for the synthetic and real cases. For the synthetic spectrum, the transitions are illustrated in Figure \ref{fig:Model_t}. The real data used is a segment of the absorption system towards HE0515-4414, $1.1494347 < z_{abs} < 1.1499145$, described in \cite{Milakovic2021}, designated ``region IV'' and illustrated in figure B4 of that paper, comprising five Fe\,{\sc ii} lines, the Mg\,{\sc ii} doublet, and one Mg\,{\sc i} line.

Two trials were carried out for both spectra, each trial having a slightly different set of first-guess parameters. In panel (a) the starting parameter guesses were different for two trials. The blue lines show the results for trial 1 and the red lines are for trial 2 (solid line = HO method, dashed line = GN-LM method). For trial 1, HO and GN-LM descend at about the same rates for the first 4 iterations, but from then on, HO performs significantly better. For the second model (red), there is little difference between the two approaches. Panel (b) shows the two trial fits to the real data. For both trials, the HO method marginally out-performs the GN-LM method although the differences are small. Interestingly however, trial 2 converges to a slightly worse $\chi^2_n$ than trial 1 but only when the GN-LM method is used; when both models are run using the HO method, the two trials converge to the same $\chi^2_n$. A visual check was made on the best-fit models and indeed the absorption component relative positions was different for the two GN-LM methods using real data. This is intriguing because it may suggest the HO method is less likely to find spurious secondary minima in $\chi^2$ space. This point must remain speculative since we have only carried out two trials here. A more detailed study should clarify this possibility. Nevertheless, the overall conclusion, albeit tentative, is that the HO method works slightly better than GN-LM.


\subsection{Stopping criteria}

Stopping criteria are discussed in detail in \cite{web:VPFIT}. The basic criterion is simply that the fitting procedure iterates until the condition $\left[F(\vx) - F(\vx+\alpha \hbox{\bf p}_{min})\right]/F(\vx) \leq \Delta$ is satisfied, where $\Delta$ can be user-defined. The appropriate value of $\Delta$ depends on the data characteristics i.e. spectral resolution, pixel size, signal to noise, and number of spectral segments being simultaneously modelled.

\section{Computing the Voigt function} \label{sec:lookuptable}

The Voigt function $H(a,u)$, a convolution of Gaussian and Lorentzian profiles. It has been described in many textbooks and papers but for the sake of completeness (and to assist other descriptive aspects of this paper), it is briefly described in Appendix \ref{apdx:Voigt}. There are published formulations of $H(a,u)$ that are derived in both frequency and wavelength space. The latter is incorrect, as discussed in Appendix \ref{app:nu_not_lambda}; the correct procedure is to use frequency space. $H(a,u)$ is expressed by an integral equation and there is a vast literature across many scientific fields describing methods for its practical computation. Numerical integration is impractical in an iterative application such as {\sc vpfit} because a huge number of repeated calculations are done. A comprehensive discussion about accuracy in computing the Voigt profile is given in \cite{MurphyPhD2002}. Whilst analytic approximations exist, the most practical approach (and the most accurate method, other than full numerical integration of the analytic Voigt function) is that introduced by \citep{Harris1948}; the Voigt function is expanded using a Taylor series and then look-up tables of the series coefficients $H_n$ as a function of $u$ used, with interpolation. This method can achieve an arbitrarily high level of accuracy, depending only on how many Taylor series terms are used, the resolution of the look-up tables, and how well interpolation in the look-up table is done (Section \ref{subsec:lookup}).

A Taylor series expansion allows the Voigt function to be expressed as
\be
\label{eq:hau}
H(a,u) = \sum_{n=0}^\infty a^n H_n(u) \,,
\ee
(see Appendix \ref{apdx:Voigt} for definitions of terms) for which the first five terms are
\be
\left.\begin{aligned} \label{eq:h0123}
H_0(u)&=e^{-u^2} \,, \\ 
H_1(u)&=-\frac{2}{\sqrt{\pi}} \left[ 1-2uD(u) \right] \,, \\
H_2(u)&=(1-2u^2) e^{-u^2}  \,, \\ 
H_3(u)&=-\frac{4}{\sqrt{\pi}} \left[ \frac{1-u^2}{3} -u \left( 1-\frac{2u^2}{3} \right) D(u) \right] \,, \\
H_4(u)&=\left( \frac{1}{2} - 2u^2 +\frac{2}{3}u^4 \right) e^{-u^2} \,,
\end{aligned}\right.
\ee
and $D(u)$ is the Dawson function,
\be
D(u) = e^{-u^2} \int_0^u e^{t^2} dt \,.
\ee

\subsection{Look-up table and interpolation} \label{subsec:lookup}

Using look-up tables in this way requires decisions that impact on the precision achieved for $H(a,u)$. First we must decide on the number of terms in the Taylor series to be included. Secondly, we must choose the number of points in each look-up table i.e. the sampling in $u$. Thirdly, we need to apply an interpolation method to extrapolate to the required value of $u$, in particular choosing the number of table points to use for polynomial fitting. To explore this, we first verified that including terms beyond $n=3$ had a negligible effect for representative values of $a$ and $u$. We then compute Eq.~\eqref{eq:hau}, summing only up to $n=3$, for combinations of look-up table resolution and the number of table interpolation points, each time calculating the fractional precision,
\be
\frac{\delta H}{H} = \left| \frac{H_{true}-H_{look-up}}{H_{true}} \right| \, .
\label{eq:deltah}
\ee
In the equation above, $H_{true}$ is calculated by full (high precision) numerical integration of the Voigt function and $H_{look-up}$ is Eq.~\eqref{eq:hau}. It is not feasible to calculate ${H_{true}}$ in real-time {\sc vpfit} usage as calculations would be too slow.

\begin{figure}
\hspace{-0.15in}\includegraphics[width=1.14\linewidth]{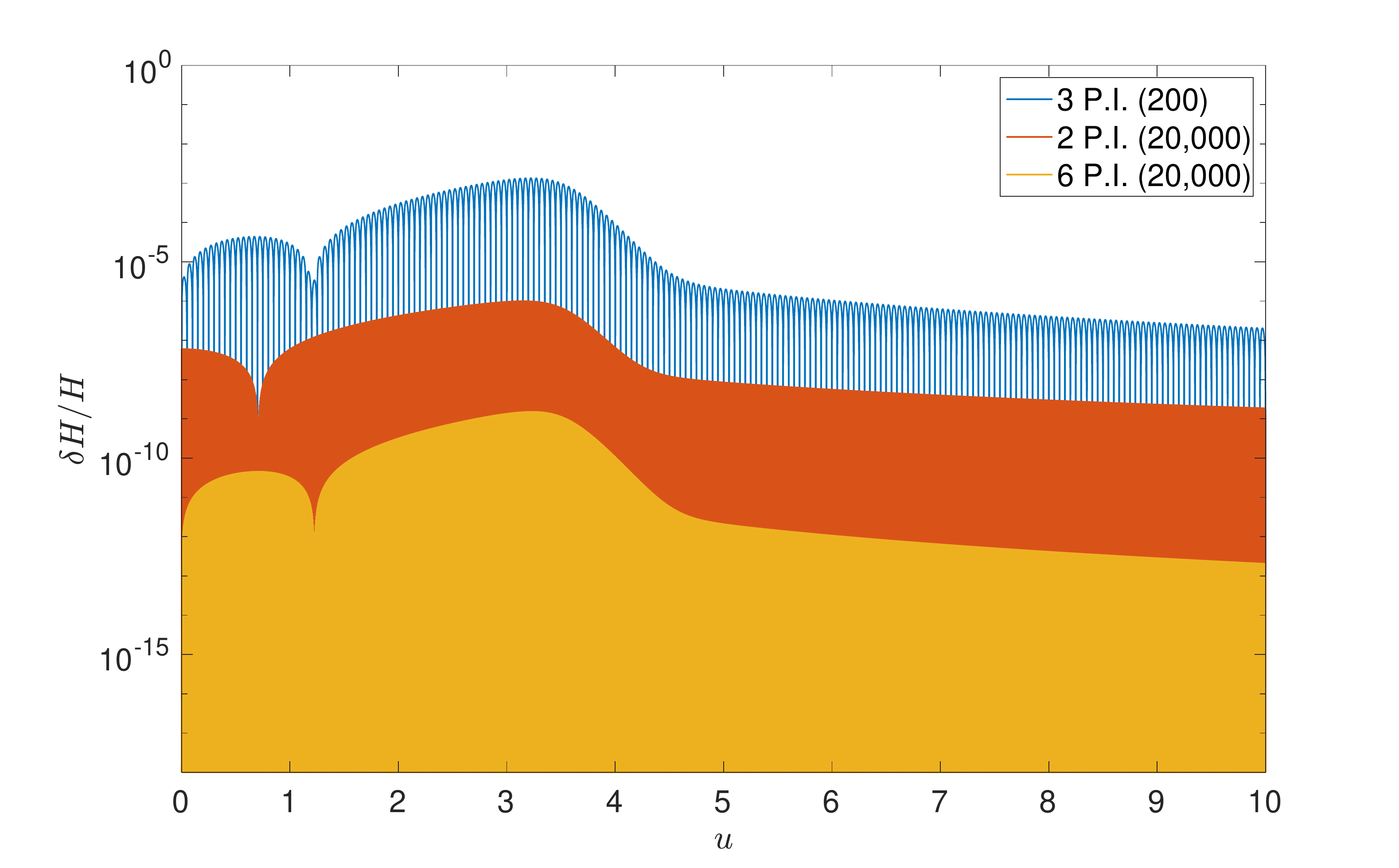}
\caption{Illustration of Eq.~\eqref{eq:deltah} for $a = 10^{-4}$ (Eq.~\eqref{eq:a}), for different resolution in the $H_n(u)$ look-up tables (Equations \eqref{eq:h0123}) and different numbers of interpolation points. The blue region illustrates $\delta H/H$ for 200 point look-up tables and 3 point interpolation. The ``line-like'' structure see in in the blue region is a consequence of the look-up table resolution; each time the look-up value of $u$ coincides exactly with an entry in the table, the precision minimises. The relative precision is worse half-way between look-up values. The red region is for 20,000 point look-up tables and 2 point interpolation. The yellow region is for 20,000 point interpolation and 6 point interpolation. The latter are the default settings adopted for use in {\sc vpfit}, so the worst-case relative precision is then $\sim 10^{-9}$. The line-like structure seen in the blue region is no longer visible in the red and yellow regions because the minima are 100x more closely spaced.}
\label{fig:interpolation}
\end{figure}

Classical Lagrange interpolation is used to extract the four coefficients $H_0(u)$, $H_1(u)$, $H_2(u)$, $H_3(u)$ for any value of the parameter $u$ (Eq.~\eqref{eq:u}). We experimented using between 2 and 6 look-up table values for interpolation. Fig.~\ref{fig:interpolation} shows the dependence of $\delta H/H$ on the number of points in the look-up table and the number of points used for interpolation within the look-up table. In practice, using 20,000 points in the look-up tables and 6-point interpolation (the default {\sc vpfit} settings adopted) yields a worst-case precision on $\delta H/H$ of $\sim 10^{-9}$, as Fig.~\ref{fig:interpolation} shows.

\section{Absorption line derivatives} \label{sec:derivs}

\subsection{Numerical derivatives -- finite differences} \label{sec:fdds}

The gradient vector and Hessian matrix in Eqs.~\eqref{eq:gradient} and \eqref{eq:hessian} need the derivative of the absorption profile in Eq.~\eqref{eq:absorption}. Finite difference derivatives ({\it fdd}) of $I(\mathbfit{x})$ generally work well although there are drawbacks. First, {\it fdd} intervals need to be assigned for every fitting parameter in the calculation. In practice this generally means one {\it fdd} interval for all column density parameters, one {\it fdd} interval for all $b$-parameters, one {\it fdd} for all $\log N$ parameters, and one {\it fdd} interval for each of the other types of free parameters. The optimal (in terms of precision) {\it fdd} interval depends on observational characteristics such as spectral resolution and spectral pixel size, as well as absorption line properties such as column density (e.g. heavily saturated lines may require different {\it fdd} intervals compared to unsaturated lines). Therefore for optimal precision, different spectral characteristics require different {\it fdd} interval settings. In principle, if using {\it fdd}, one could calculate optimal {\it fdd} intervals that minimise the overall error on the target function. Methods for doing this are described in \cite{GMW81} (section 4.6.1.3) and in \cite{Press2007}. Nevertheless, in practice it is still difficult to assign individual intervals for {\it every} parameter for every velocity component in an absorption complex. This problem can be completely eliminated by replacing the {\it fdd} with analytic derivatives. We therefore next describe an efficient way of calculating analytic derivatives of the Voigt function. The new method builds on the widely used Taylor series expansion method applied to the Voigt function itself \citep{Harris1948}. Surprisingly, as far as we know, the idea has not been applied  to {\it derivatives} of the Voigt function before (or indeed to other parameter derivatives relevant to absorption line spectroscopy).

\subsection{Voigt function and other analytic derivatives}
\label{sec:analyticderivs}

Derivatives of $I(\mathbfit{x})$ can be expressed in terms of derivatives of $H(a,u)$. The \cite{Harris1948} method can also be used to calculate analytic derivatives, as we now show.
\be
\frac{dI_\nu}{dx_i} = I_\nu \left( \frac{1}{I_{\textrm{o}}} \frac{dI_{\textrm{o}}}{dx_i} -\sum_{j=1}^m \frac{d \tau_j}{d x_i} \right) \,,
\label{eq:dIdx}
\ee
where $I_{\textrm{o}}$ is the unabsorbed continuum intensity and for simplicity the explicit frequency dependence on $\tau$ (subscript) has been dropped so that we can assign an index for each absorption component. Ignoring convolution with the instrumental profile for the moment (until Section \ref{sec:derivconvolve}), this function can be calculated analytically, making use of Eqs.~\eqref{eq:tau} and \eqref{eq:A3},
\be
\frac{d \tau_j}{d x_i} = \bar{\kappa}_j \frac{d N_j}{dx_i} + N_j \frac{d \bar{\kappa}_j}{dx_i} \,.
\ee
For each fitting parameter, $x_i$, analytic derivatives can be specified as follows.\\

\noindent {\bf (i) Redshift ($x_i = z_i$):}
\be
\left.\begin{aligned}
\label{eq:diff_z}
&d I_{\textrm{o}} / dz_i = dN_j / dz_i =0 \,, \\
&\frac{da_j}{dz_i} = 0 \,, \\
&\frac{du_j}{dz_i} = \frac{c}{b} \frac{\lambda_o}{\lambda}\,, \\
&\frac{d \tau_j}{dz_i} = - \frac{\sqrt{\pi}e^{2}f N_i H_{,u}}{m_{e} b_i \Delta \! \nu_{d}} \frac{\lambda_o}{\lambda} \delta_{ij} \,,
\end{aligned}\right.
\ee
where $\lambda_o$ is the rest-frame atomic wavelength, $\delta_{ij}$ is the Dirac delta function and $H_{,u} = \partial H/ \partial u$.

\vspace{10pt}
\noindent {\bf (ii) $b$-parameter ($x_i = b_i$):}
\be
\left.\begin{aligned}
\label{eq:diff_b}
&d I_{\textrm{o}} / db_i = dN_j / db_i = 0 \,, \\
&\frac{da_j}{db_i} = -\frac{a_i}{b_i} \delta_{ij} \,, \\ 
&\frac{du_j}{db_i} = -\frac{u_i}{b_i} \delta_{ij} \,, \\
&\frac{d \tau_j}{db_i} = - \frac{\sqrt{\pi}e^{2}f N_i}{m_{e}c\Delta \! \nu_{d}} \left( \frac{H}{b_i} + \frac{u_i H_{,u}}{b_i} + \frac{a_i H_{,a}}{b_i} \right) \delta_{ij} \,,
\end{aligned}\right.
\ee
where $H_{,a} = \partial H/ \partial a$.

\vspace{10pt}
\noindent {\bf (iii) Column density ($x_i = \log \! N_i$):}
\be
\left.\begin{aligned}
\label{eq:diff_N}
&\frac{d I_{\textrm{o}}}{d \log \! N_i} = \frac{du_j}{d \log \! N_i} = \frac{da_j}{d \log \! N_i} =0 \,, \\
&\frac{dN_j}{d \log \! N_i} = ( N_i \ln\! 10 ) \delta_{ij} \,, \\
&\frac{d \tau_j}{d \log \! N_i} = (\tau_i \ln\! 10) \delta_{ij} \,.
\end{aligned} \right.
\ee

To solve equations \eqref{eq:diff_z} to \eqref{eq:diff_N}, the Voigt function derivatives $H_{,a}$ and $H_{,u}$ are needed, which are
\be
H_{,a} = \sum_{n=1}^\infty n a^{n-1} H_n
\quad \mathrm{and} \quad H_{,u} = \sum_{n=0}^\infty a^n H_{n,u} \,,
\ee
where we have dropped notating the $(a,u)$ dependence for simplicity. It is usually true that $a \ll 10^{-2}$, so that we can easily reach an accuracy of $\ll 10^{-10}$ if we expand to fourth order. The first four terms of $H_{,u}$ can be obtained from
\be
\left.\begin{aligned}
&H_{0,u} = -2u e^{-u^2} \,, \\
&H_{1,u} = \frac{4}{\sqrt{\pi}}\left[ u+ \left(1-2u^2 \right)D(u) \right] \,, \\
&H_{2,u} = \left(-6u + 4u^3\right)e^{-u^2} \,, \\
&H_{3,u} = \frac{4}{3\sqrt{\pi}}\left[ \left(5u - 2u^3 \right) + \left( 3-12u^2+4u^4 \right) D(u) \right] \,,
\end{aligned} \right.
\ee
where we have used
\be
\frac{dD(u)}{du} = 1- 2u D(u) \,.
\ee

We adopt, as default internal settings in {\sc vpfit}, the same set of parameters for the resolution of the four look-up tables (20,000 points in each) and 6-point interpolation as before.

\vspace{10pt}
\noindent {\bf (iv) Continuum level:}

The continuum level for each spectral segment in the absorption system model can be varied within {\sc vpfit}. This is done using a two-parameter fit, the constant normalisation $I_{\textrm{o}}^{(0)}$ and the first order derivative of the existing (i.e. the user-provided) continuum level $I_{\textrm{o}}^{(1)}$,
\be
I_{\textrm{o}} = I_{\textrm{o}}^{(0)} + I_{\textrm{o}}^{(1)} (\lambda - \lambda_c) \,,
\ee
where $I_{\textrm{o}}^{(0)} = I_{\textrm{o}}\rvert_{\lambda = \lambda_c}$, $I_{\textrm{o}}^{(1)}= \left( d I_{\textrm{o}}/d\lambda \right) \rvert_{\lambda = \lambda_c}$ and $\lambda_c$ is a user-provided wavelength (commonly selected as the centre of the spectral segment being fitted). The analytical derivatives of both terms are
\be
\left.\begin{aligned}
& \frac{d I_\nu}{d I_{\textrm{o}}^{(0)}} = \frac{ I_\nu }{I_{\textrm{o}}} \,, \\
& \frac{d I_\nu}{d I_{\textrm{o}}^{(1)}} = (\lambda - \lambda_c) \frac{  I_\nu }{I_{\textrm{o}}} \,.
\end{aligned}\right.
\label{eq:continderivs}
\ee
Equations \eqref{eq:continderivs} are used to calculate the Hessian matrix and gradient vector components such that the two free continuum parameters may be solved for in the same way as the absorption line Voigt parameters.

\vspace{10pt}
\noindent {\bf (v) Zero level adjustment:}

During the spectral data reduction, it may sometimes be that sky subtraction is imperfect, or that some scattered light in the spectrograph causes a residual, non-astrophysical zero level offset in the spectrum being analysed. There may be other sources of scattered light. If this is not properly accounted for or removed prior to Voigt profile modelling, the Voigt parameters will be systematically biased. In practice, a continuum may be fitted {\it prior to} any attempt to detect any residual background in the observed spectrum. To properly account for this, we can model the observed spectrum as
\be
\left.\begin{aligned}
& I_\nu = Z_{\textrm{o}} + (1-Z_{\textrm{o}}) I_{\textrm{o}} \exp \left( - \sum_{j=1}^m \tau_j \right) \,, \\
& \frac{d I_\nu}{d Z_o} = 1 - I_{\textrm{o}} \exp \left( - \sum_{j=1}^m \tau_j \right) \,.
\end{aligned}\right.
\label{eq:zero}
\ee
Expressed in this way, the parameter $Z_o$ allows for a residual background and corrects the local continuum level appropriately. The derivative in Eq.~\eqref{eq:zero} is then used to form the required terms in the gradient vector and Hessian matrix.

\vspace{10pt}
\noindent {\bf (vi) Velocity shift:}

It is useful to have the capability of including an additional free parameter to each spectral segment that allows a shift (in velocity space) relative to other spectral segments. This may be desirable for a number of reasons, for either astrophysical reasons or to allow for for possible systematic uncertainties in the zero point of the wavelength calibration. This can be done simply, as follows.

The velocity shift effectively contributes to the redshift parameter of all absorption components. This can be expressed in terms of the distance from the line centre in Doppler width units, $u$,
\be
\left.\begin{aligned}
& \frac{du_j}{d v} = \sum_i \frac{du_j}{dz_i} \frac{dz_i}{dv} = \sum_i \left( \frac{1+z_i}{c} \frac{du_j}{dz_i} \right) \,, \\
& \frac{d\tau_j}{d v} = \sum_i \frac{d\tau_j}{d z_i} \frac{dz_i}{d v} = \sum_i \left(\frac{1+z_i}{c} \frac{d\tau_j}{dz_i} \right) \,.
\end{aligned}\right.
\label{eq:shifts}
\ee
where $v$ is the velocity shift parameter and $du_j/dz_i$, $d\tau_j/dz_i$ follow Eq. \eqref{eq:diff_z}. Equations \eqref{eq:shifts} are used to calculate the derivatives required by the Hessian matrix and gradient vector.

\subsection{Comparing finite difference and analytic derivatives} \label{sec:analy_vs_fdd}

It is informative to compare the relative precision of the {\it fdd} and Taylor series derivative methods for different absorption line parameters. To do so we examine derivatives with respect to the $b$-parameter, for two cases, unsaturated and saturated absorption lines, defining the following quantities:
\be
\left.\begin{aligned}
\label{eq:dIdb}
I_{\nu,b} = \frac{d I_\nu}{db} \,,
\end{aligned} \right.
\ee
\be
I_{\nu,b}^{\it fdd} = \frac{I_\nu \rvert_{b+\delta b} - I_\nu\rvert_{b-\delta b}}{2 \delta b} \,,
\ee
\be
\left.\begin{aligned}
I_{\nu,b}^{\rm look-up} &= \frac{d I_\nu}{db} = I_{\textrm{o}} \frac{d e^{-\tau_\nu}}{db} = - I_\nu \frac{d \tau_\nu}{db} \\ 
& = \frac{I_\nu \tau_\nu}{b} \left( 1 + \frac{u H_{,u}}{H} + \frac{aH_{,a}}{H} \right)\,, 
\end{aligned} \right.
\ee
where $H$, $H_{,a}$ and $H_{,u}$ are calculated using look-up tables as described in Section \ref{sec:analyticderivs}. We note again for clarity that $I_\nu$ is the spectrum prior to convolution with the instrumental profile, whereas $I(\mathbfit{x})$ is the convolved spectrum, $(I_{\nu} \ast \mathcal{G})$.

\begin{figure*}
\centering
\includegraphics[width=0.49\linewidth]{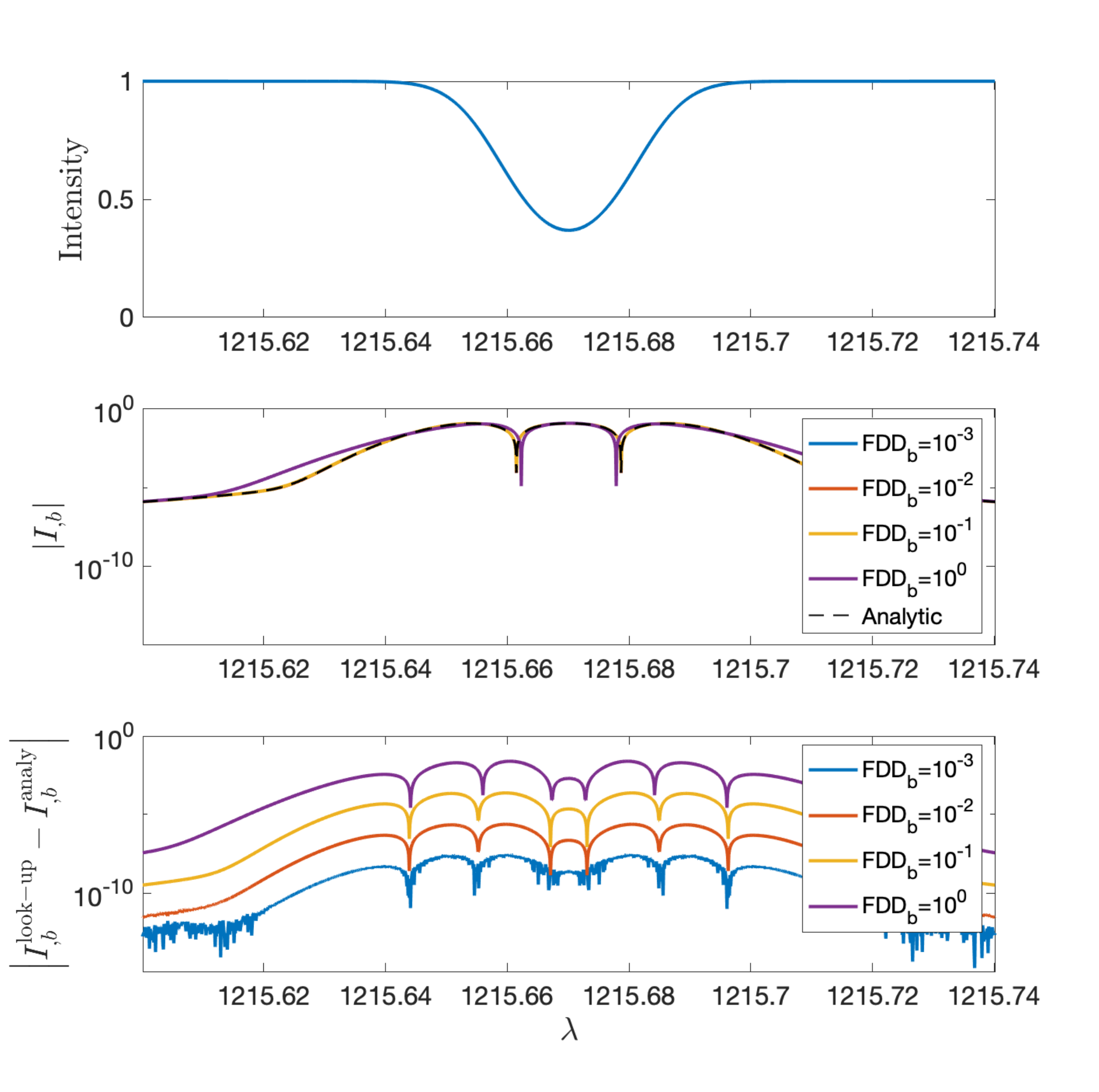}
\includegraphics[width=0.49\linewidth]{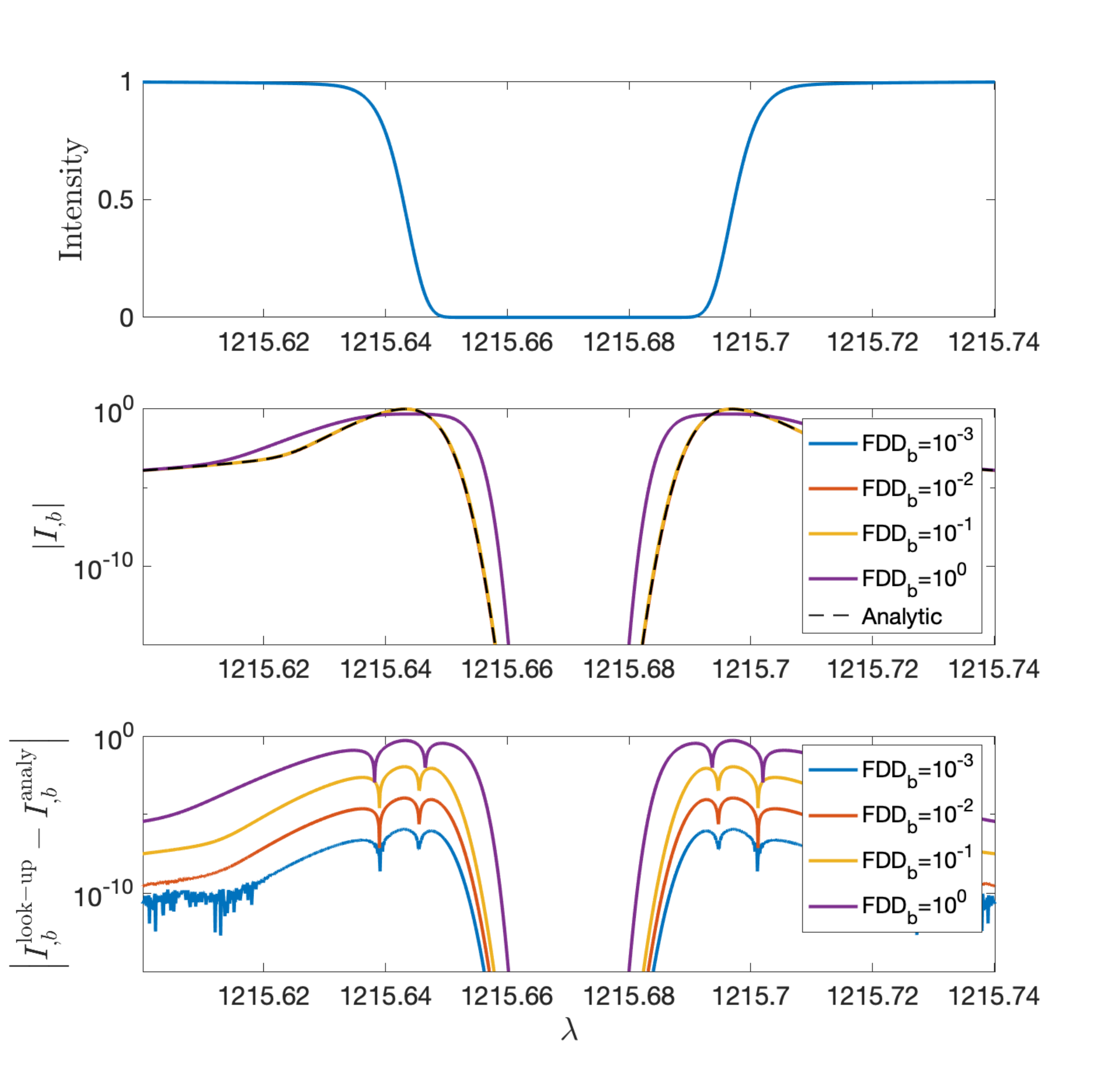}
\includegraphics[width=0.49\linewidth]{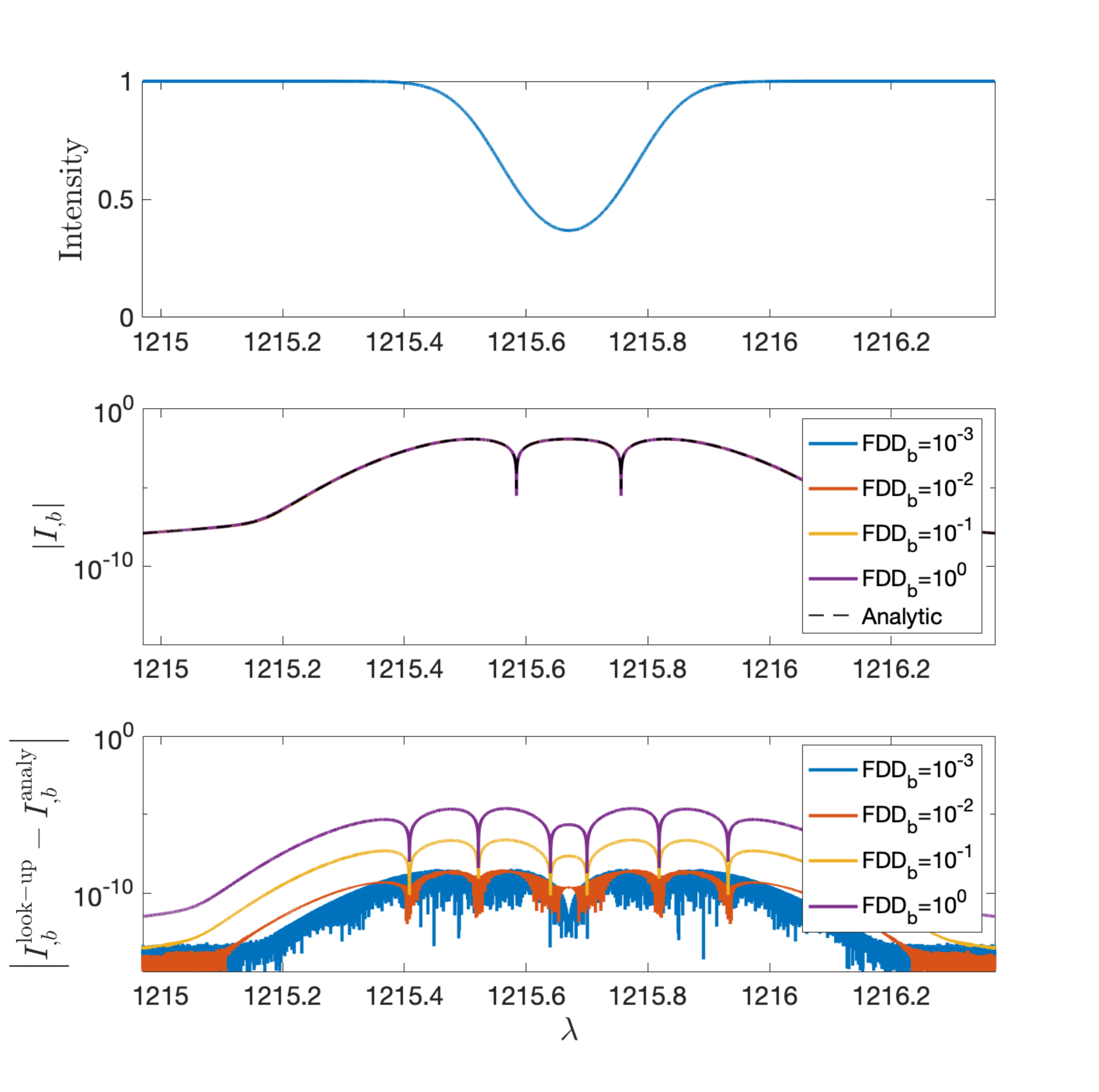}
\includegraphics[width=0.49\linewidth]{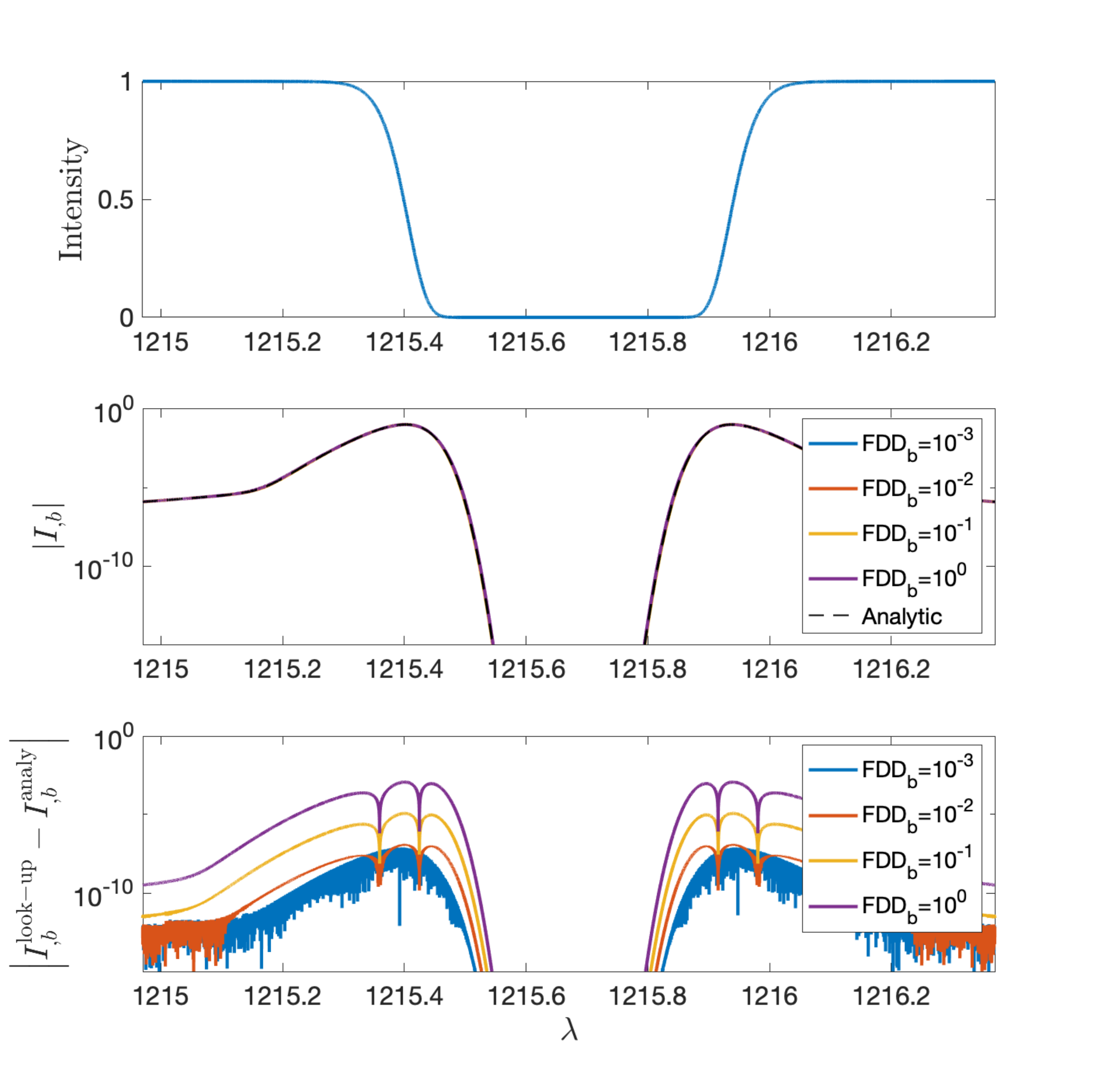}
\caption{Plots of $I(\lambda)$, $I_{,b} = dI/db$ (Eq.~\eqref{eq:dIdb}), and the difference between {\it fdd} and analytic derivatives, $I_{,b}^\mathrm{look-up} - I_{,b}^\mathrm{analytic}$ for a hydrogen Lyman-$\alpha$ absorption line. See Section \ref{sec:analy_vs_fdd} for details.}
\label{fig:analy_numer}
\end{figure*}

In Fig.~\ref{fig:analy_numer}, four sets of three panels are illustrated. The top 6 panels are for $b=3$ and the lower 6 panels are for $b=30$ km/s. The range in $b$-parameters covers the majority of observed parameters for both hydrogen and heavy element lines. Whilst $b=3$ is unrepresentative of typical hydrogen lines seen in quasar spectra, it is not atypical for heavy element lines so we represent the calculations this way, using HI as the atomic transition, for convenience (so that other quantities, notably $\Gamma$ and the central wavelength, remain constant).

The optical depth is given by
\be
\tau_\nu = N \frac{\sqrt{\pi} e^2}{m_e c} \frac{f_\lambda}{\Delta \nu_d} H = \kappa_0 N H \,.
\ee
See Appendix \ref{apdx:Voigt} for definition of quantities. The panels on the left hand sides of Fig.~\ref{fig:analy_numer} are for an unsaturated absorption line with $\kappa_0 N = 1$ ($b=3$ km s$^{-1}$) and  $\kappa_0 N = 10$ ($b=30$ km s$^{-1}$), corresponding to a column density of $\log N = 12.55$ and $\log N = 13.55$ atoms cm$^{-2}$. The right hand panels are for a saturated line with  $\kappa_0 N = 100$ ($b=3$ km s$^{-1}$) and  $\kappa_0 N = 1,000$ ($b=30$ km s$^{-1}$), corresponding to a column density of  $\log N = 14.55$ and $\log N = 15.55$ atoms cm$^{-2}$. Each grouping of 3 panels plots the intensity $I$, the analytic derivative with respect to the velocity dispersion parameter $| I_{,b}^{look-up} |$, and the absolute difference between the analytic and {\it fdd} derivatives $| I_{,b}^{look-up} - I_{,b}^{\it fdd} |$, as a function of rest-frame wavelength. The derivative panels illustrate several models, discussed in the figure caption and indicated by the figure legends.

The main things we learn from the quantities illustrated in Fig.~\ref{fig:analy_numer} are: \\
1. For the range of absorption line parameters considered, {\it fdd} and analytic derivatives agree well for suitably small finite difference intervals,\\
2. However, for unsuitably large {\it fdd} increments (e.g. 0.1 km/s or larger) the absolute difference between {\it fdd} and analytic derivatives can be as large as $\sim 10^{-4}$ for unsaturated lines and $\sim 10^{-2}$ for saturated lines, corresponding to percentage differences of $\sim 0.1$ and $\sim 1$\% respectively.

Translating the {\it fdd} precision constraints into practical quantities such as non-linear least squares descent efficiency is complicated because there are generally many variables involved. In the idealised case considered here, there is only one single absorption line. In any real quasar absorption system, there are multiple blended components. In a complex absorption system (comprising multiple blends), some absorption parameters may be well determined whilst others may be very poorly determined or even completely degenerate. In these cases, one single {\it fdd} interval setting may be optimal for some absorption components, but inappropriate for others. Analytic derivatives do not suffer from this difficulty and hence completely avoid the potential problems arising through choice of {\it fdd} interval.

\subsection{Convolution with the instrumental profile for analytic derivatives} \label{sec:derivconvolve}

When {\it fdd} are calculated from the model profiles, convolution with the instrumental profile has already taken place, so the instrumental resolution is properly accounted for. However, the analytic derivatives of Section \ref{sec:analyticderivs} have ignored the instrumental profile. The derivative of a convolution theorem e.g. \citet{Bracewell1978} can be employed,
\be
\frac{\partial \left( I_{\nu} \ast \mathcal{G} \right)}{\partial x_i} = \frac{\partial I_{\nu}}{\partial x_i} \ast \mathcal{G} \,,
\label{eq:derivconvolve}
\ee
where $\ast$ denotes convolution, $x_i$ is the model parameter used in Sec.~\ref{sec:analyticderivs}, and $\mathcal{G}$ is the instrumental resolution. The required analytic derivatives, i.e. the left hand side of Eq.~\eqref{eq:derivconvolve}, are thus obtained by first calculating Eqs.~\eqref{eq:dIdx} and then convolving those derivative arrays with the appropriate function $\mathcal{G}$ (e.g. a Gaussian or a numerical function).

\section{Discussion} \label{sec:discussion}

New astronomical facilities such as the ESPRESSO spectrograph on the European Southern Observatory’s VLT \citep{Pepe2014} and the forthcoming ELT, will carry out large explorations for new physics \citep{Hook2009, ESO_ELTbook2010, ESO_ELTbook2011, ELT_Liske2014, Marconi2016}. Two areas of particular study will be searches for temporal or spatial variations in fundamental constants and measurements of cosmological redshift drift. Both of these projects require maximally precise absorption line modelling procedures. These things motivate the analytic and numerical methods presented in this paper. Focusing on precision and computational practicality, we have aimed primarily to do the following:
\begin{enumerate}
    \item present a detailed account of the theoretical methods on which the widely used code {\sc vpfit} is based,
    \item introduce a fast and precise method for calculating Voigt function derivatives, and other relevant derivatives, newly implemented in {\sc vpfit},
    \item describe a new optimisation method, incorporating the simultaneous use of two different fine tuning methods drawn from the Gauss-Newton and Levenberg-Marquardt approaches.
\end{enumerate}

\section*{Acknowledgments}

We are grateful to Mike Irwin for help with an early implementation of the Gauss-Newton method to Voigt profile modelling. We thank Jochen Liske for pointing out the importance of sufficiently high sampling of the Voigt profile. Several people contributed either by making {\sc vpfit} code modifications or by testing, including Andrew Cooke, Vincent Dumont, Matthew Bainbridge, Julian King, Dinko Milakovi{\'c}, and Michael Murphy. We thank Pasquier Noterdaeme and C\'edric Ledoux for raising the question of frequency vs. wavelength formulation for the Voigt function. JKW thanks the John Templeton Foundation for support.

\section*{Data Availability}
The {\sc vpfit} code and data used in this paper is available at \url{https://people.ast.cam.ac.uk/~rfc/}. Any additional material not included there can be requested directly from the authors.

\bibliography{vpfit}
\bibliographystyle{mnras}

\appendix
\section{ABSORPTION PROFILE}

\subsection{Voigt profile} \label{apdx:Voigt}

For a complex of $m$ blended absorption profiles, the intensity as a function of frequency is given by 
\begin{equation}
\label{eq:absorption}
I_{\nu} = I_{\textrm{o}} \exp \left( - \sum_{j=1}^m \tau_j \right),
\end{equation}
where $I_{\nu}$ is the observed intensity, $I_{\textrm{o}}$ is the unabsorbed continuum intensity, and $\tau_j$ is the optical depth. Instrumental resolution is not included here. It is assumed (in {\sc vpfit}) that a single opacity applies to a single absorption component in a complex, i.e. that 
\begin{equation}
\tau_j = \bar{\kappa}_j N_j \, ,
\label{eq:tau}
\end{equation}
where $N$ is the column density of absorbing atoms and the absorption coefficient (averaged over the gas cloud) is 
\begin{equation}
\label{eq:A3}
\bar{\kappa}_j = \frac{\sqrt{\pi}e^{2}f_jH(a,u)_j}{m_{e}c\Delta \! \nu_{d}}
\end{equation}
e.g. \citet{Mihalas1978}. Hereafter, we focus on a single absorption component and drop the subscript $j$. $H(a,u)$ is the Voigt function, $f$ is the atomic oscillator strength, $e$ and $m_e$ are the electron charge and mass, and $c$ is the speed of light. The Doppler width $\Delta \! \nu_{d}$ is related to the velocity dispersion parameter $b$ and the line central frequency in the rest frame $\nu_{o}$ by 
\begin{equation}
\Delta \! \nu_{d} = b\nu_{o}/c \, .
\label{eq:dopplerwidth}
\end{equation}
In general, the observed velocity dispersion parameter is the quadrature addition of the thermal and turbulent broadening \citep{Struve1934},
\begin{align}
b_{\textrm{obs}}^2 &= b_{\textrm{thermal}}^2 + b_{\textrm{turbulent}}^2 \nonumber \\
&= \frac{2kT}{M} + b_{\textrm{turbulent}}^2 \, ,
\end{align}
where $M$ is the atomic mass.

The Voigt function $H(a,u)$ is the ratio of the (frequency dependent) absorption coefficient to the absorption coefficient at the line centre, i.e.
\be
 H(a,u) = \frac{\bar{\kappa}_{\nu}}{\kappa_0} \, ,
\ee
where
\be
\label{eq:kappa0}
\kappa_0 = \frac{\sqrt{\pi}e^{2}f}{m_{e}c\Delta \! \nu_{d}} \, .
\ee
$H(a,u)$ is expressed in terms of two dimensionless parameters. The first, 
\be
a=\Gamma/4\pi \Delta \! \nu_{d} \, ,
\label{eq:a}
\ee
gives the ratio of the natural line width to the Doppler width and  depends on the damping constant, the frequency of the line centre, the temperature, and the atomic mass. $\Gamma$ is the damping constant, the sum of the spontaneous emission rates. If the transition is from the ground state, this is
\begin{equation}
\Gamma = \sum_{j=1}^k A_{kj} \, ,
\label{eq:gamma1}
\end{equation}
where $k$ refers to the upper level. If the lower level is also an excited state, 
\begin{equation}
\Gamma = \sum_{i=1}^l \Gamma_{li} + \sum_{i=1}^u  \Gamma_{ui} \, .
\label{eq:gamma2}
\end{equation}

The second dimensionless argument of the Voigt function expresses the distance from the line centre in Doppler width units,
\be
u = \frac{\nu_r-\nu_{o}}{\Delta \! \nu_{d}} = \frac{\lambda_r-\lambda_{o}}{b \lambda_r / c} \, ,
\label{eq:u}
\ee
where $\lambda_{o} = c/\nu_{o}$, $\nu_r = \nu (1+z)$ and $\lambda_r=\lambda/(1+z)$ are the rest-frame frequency and wavelength, $\lambda$ is the observed-frame wavelength, and $z$ is redshift. Since $\Delta \! \nu_{d}$ is constant, $u$ is symmetric in frequency space. Optical/UV spectra are generally plotted (and analysed) in wavelength space, which is what {\sc vpfit} uses.

\subsection{Should the Voigt function be formed in frequency space or in wavelength space?} \label{app:nu_not_lambda}

This seemingly trivial question is in fact important\footnote{Nikola Tesla reputedly said: ``If you want to find the secrets of the Universe, think in terms of energy, frequency and vibration.''}. Historically, the semi-classical Voigt function was derived in frequency space. {\sc vpfit} uses frequency space (Eqs.~\eqref{eq:dopplerwidth} and \eqref{eq:u}). However, formulations of the Voigt profile exist in which wavelength rather than frequency is the function variable. Examples of this are \cite{Whiting1968}, the analytic approximation of \cite{Garcia2006}, codes such as FITLYMAN \citep{Fontana1995}, and VoigtFit \citep{Krogager2018}. Given the broad usage of the Voigt function, there may be many other codes that also use wavelength space, particularly in fields other than astrophysics.

A Voigt function in which the $u$ parameter is defined as 
\be
u = \frac{\lambda_r-\lambda_{o}}{b \lambda_{o} / c}
\label{eq:wrongu}
\ee
cannot share the same symmetry as a Voigt function using Eq.~\eqref{eq:u}. The symmetry difference increases with increasing distance from the line centre as can be seen in the two damped Lyman $\alpha$ (DLA) Voigt profiles  illustrated in Fig.~\ref{fig:3DLAs}. The red dashed line (using Eq.~\eqref{eq:u}) is symmetric in frequency space and the orange dotted line (using Eq.~\ref{eq:wrongu}) is symmetric in wavelength space. Both cannot correctly represent Nature and the following examples suggest that frequency is the fundamental quantity:
\begin{enumerate}
\item The energy level spacings in an atom define transition frequencies. The natural line width of a transition is given by the lifetimes of the upper and lower states, as required by the Heisenberg Uncertainty principle, thus is defined by the damping constant $\Gamma$ in Hertz (Eqs.\ref{eq:gamma1} and \ref{eq:gamma2}).
\item Wavelength, not frequency, changes when light crosses a refractive index boundary. This is obvious from de Broglie's wave equation, $\lambda = h/p$; as a photon's momentum changes, so does its wavelength.
\item Noether’s Theorem shows that invariance under time translation leads to the principle of conservation of energy. Changing the frequency in the same reference frame would violate energy conservation.
\end{enumerate}
These considerations illustrate that frequency is the more fundamental quantity and hence that the correct formulation of $H(a,u)$ should be in frequency space; Eq.~\eqref{eq:u} should always be used. Eq.~\eqref{eq:wrongu} should not be used.

Searches for spacetime variations of fundamental constants require placing constraints on tiny shifts in spectroscopic transitions. At lower column densities (relevant for high redshift varying fine structure constant measurements using heavy element absorption lines), the symmetry difference between the $u_{\nu}$ and $u_{\lambda}$ profiles is small although it increases with column density. This can be illustrated by comparing line centroids,
\be
\mathrm{Centroid_{H\nu, H\lambda}} = \frac{\int_{-\infty}^{\infty} \lambda \left( 1-I_{H\nu, H\lambda} \right) d \lambda}{\int_{-\infty}^{\infty} \left( 1-I_{H\nu, H\lambda} \right) d \lambda} \, ,
\ee
where $I_{H\nu, H\lambda}$ represents the model intensity calculated either using the Voigt function defined in frequency space, or the Voigt function defined in wavelength space. The centroid shift between Lyman $\alpha$ lines calculated both ways is about $1.3 \times 10^{-5}${\AA} for $N_{HI} = 10^{14}$ atoms cm$^{-2}$, increasing to $0.11${\AA} for $N_{HI} = 10^{20}$ atoms cm$^{-2}$ (both computed using a representative $b$-parameter of 15 km s$^{-1}$, although these shifts are relatively insensitive to $b$). A centroid shift of $1.3 \times 10^{-5}$~{\AA}, assuming $z_{abs}=3$, means that Eq.~\eqref{eq:wrongu} gives a line position that is wrong by $\sim 1$~m/s. The present best-case fractional uncertainty on the measured redshift of a heavy element absorption line is $\sim 1 \times 10^{-7}$, which is 30 times larger. Nevertheless, Eq.~\eqref{eq:wrongu} should be avoided since systematic biases could accumulate over large statistical samples of high signal to noise measurements.\\

\subsection{The Kramers-Heisenberg-Thermal (KHT) profile} \label{sec:A_KHT}

The Voigt function treats each atomic transition in isolation as a classical 2-level damped harmonic oscillator, ignoring interactions associated with other discrete levels, and ignoring transitions from continuum energy states to other bound levels. Despite this, the Voigt model is very accurate except at higher column densities, where interactions between multiple discrete levels and with continuum transitions become important.

Deficiencies of the Voigt profile in some applications have been widely discussed. \cite{Tennyson2014} discuss an alternative to the Voigt profile, the Hartmann–Tran profile, which some authors recommend for high resolution spectroscopy, eg. \cite{Schreier2017}. \cite{Lee2013} and earlier papers by the same authors have highlighted significant departures between the semi-classical Voigt profile and the quantum mechanical Kramers-Heisenberg profile. However, unlike the Voigt profile, the Kramers-Heisenberg profile does not include thermal broadening, prompting \cite{Lee2020KH} to develop the Kramers-Heisenberg-Thermal profile (KHT). One cannot, of course, define an exact column density below which Voigt is accurate and above which KHT should be used, as this depends on data quality and the required precision. However, as a rule-of-thumb, one should be wary of applying the Voigt profile to column densities above $\sim 10^{20}$ atoms cm$^{-2}$. 

\begin{figure}
\centering
\includegraphics[width=1.05\linewidth]{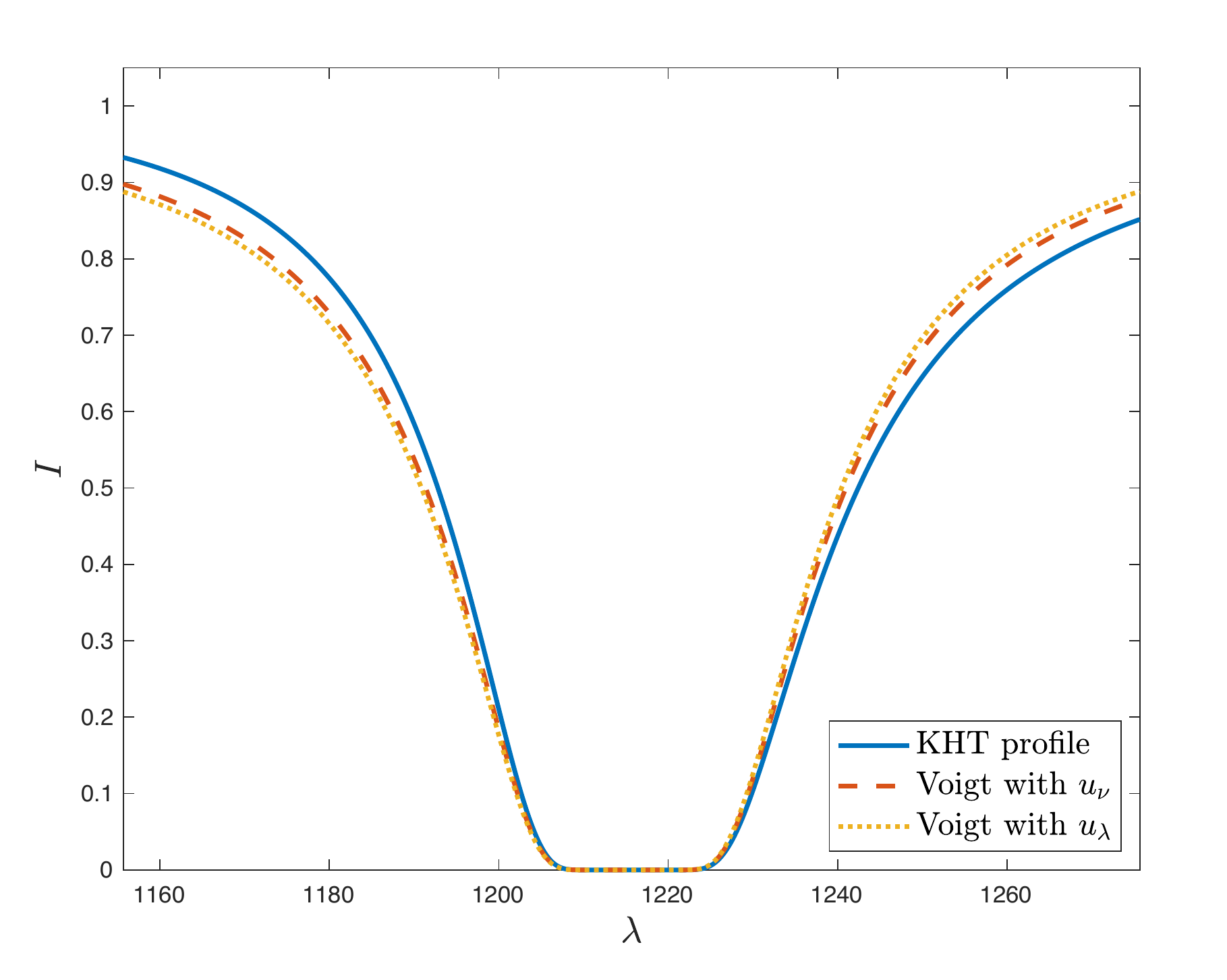}
\caption{Damped Lyman-$\alpha$ profile with $\log N_{HI} = 22.0$ and $b=15$ km\,s$^{-1}$. Blue continuous line: Kramers-Heisenberg-Thermal profile \citep{Lee2020KH}. Red dashed profile: Voigt profile using $u_{\nu}=(\nu-\nu_{o})/\Delta \! \nu_{d}$ and $\Delta \! \nu_{d} = b\nu_{o}/c$. Orange dotted line: Voigt profile using 
$u_{\lambda}=(\lambda_{o}-\lambda)/ (b \lambda_{o}/c)$.
}
\label{fig:3DLAs}
\end{figure}

Fig.~\ref{fig:3DLAs} provides a comparison between the Voigt profiles (derived in frequency and wavelength space) and the KHT profile, for a column density of $10^{22}$ atoms cm$^{-2}$. The KHT profile has been implemented in {\sc vpfit}v12.2. \\

\section{FURTHER NUMERICAL CONSIDERATIONS}

\subsection{The importance of sub-binning in practical calculation}

When computing Eq.~\eqref{eq:absorption}, it is essential to do so using a fine grid, usually much finer than the original spectral data. The sub-bin pixel size must be small enough for the shape of the absorption profile to be effectively linear over the wavelength range spanned by a single sub-pixel. If this is not done (i.e. if instead the profile is simply evaluated at the centre of a large pixel), the opacity model will be wrong. In practice, {\sc vpfit} assigns a default value but this can be user-defined if required.

\subsection{Parameter update bounds}

When the model parameters are far from the best-fit solution, the parameter update vector $\mathbfit{p}$ may be unstable such that one or more parameters are driven even further away from the correct solution. The simple solution to this problem is to constrain each parameter update to be within some upper bound. Empirically, the most unstable parameter is probably $\log N$. When bad blending/ill-conditioning occurs, this parameter can be poorly constrained and, if left unbound, can shoot to very low or implausibly high values. The problem can be avoided i.e. stability can be improved, by limiting the maximum step at any {\sc vpfit} iteration. Similar effects can occur for $b$-parameter and $\daa$. Parameter ties amongst several species \citep{web:VPFIT} helps to avoid these problems (for $b$ and of course for $\daa$), but not always. Overall parameter bounds are already included in {\sc vpfit}v12.3. As of {\sc vpfit}v12.3 maximum step sizes (within each {\sc vpfit} iteration) can also be user-defined, or default presets can be used.

\subsection{Sensitivity of best-fit model parameters to initial parameter guesses} \label{sec:firstguesses}

For one single-component unblended absorption system, provided stopping criteria are appropriately set, there should be no sensitivity of the final best-fit model to initial parameter guesses. However, for multiply blended absorption components, this is not the case and in general it is possible to find multiple models that give similarly statistically acceptable representations of the data.

{\sc ai-vpfit}, is an artificial intelligence version of {\sc vpfit} that builds on the work of \cite{gvpfit2017, Bainbridge2017} and is described in \cite{Lee2020AI}. It employs a Monte Carlo method for generating first guess parameters for each absorption component fitted in an absorption complex. Independent runs of {\sc ai-vpfit} on the same spectrum use different random number seeds for trial absorption component placement. This means that model construction progresses differently in each {\sc ai-vpfit} run, emulating the different approaches that would be taken by different humans doing the job interactively using {\sc vpfit}. By running {\sc ai-vpfit} multiple times, \cite{Lee2021} explored the sensitivity of fine structure constant ($\alpha$) measurements to the initial parameters guesses; multiple {\sc ai-vpfit} runs end up with slightly different final answers, revealing sub-structure in $\chi^2$-parameter space. In other words, multiple {\sc ai-vpfit} runs reveal model non-uniqueness. \cite{Lee2021} show that the final best fit values of $\alpha$ may indeed depend on the manner in which models are constructed. 

The degree of model non-uniqueness was shown to depend on the particular Information Criterion used (the corrected Akaike  Information criterion (AICc), the Bayesian Information Criterion (BIC), or the Spectroscopic Information Criterion (SpIC)), as well as the physical line broadening mechanism used (turbulent, thermal, or compound broadening). Since only two high redshift quasar absorption systems were studied in \cite{Lee2021}, the extent to which these things are generally true awaits a larger study.

In the context of {\sc vpfit} usage, it may be assumed that the best-fit model parameters could depend on the way in which the model for an absorption complex comprising multiple components is constructed.

\subsection{Information criteria; avoiding over- or under-fitting} \label{sec:ICs}

Over-fitting with too many model parameters creates spurious line bends, in turn causing potentially biased results and unnecessarily increased parameter errors. Under-fitting results in artificially small parameter uncertainties but increased scatter \citep{Wilczynska2015}. Using the value of best-fit $\chi^2$ as an indicator of the appropriate number of free parameters to use for any model is unreliable for two reasons: (a) $\chi^2$ does not minimise as a function of the number of free parameters so only a lower limit on the number of free parameters can be obtained, and (b) in any case, it is generally the case that the spectral error array is imperfect such that the absolute value of $\chi^2$ is uncertain. For these reasons, the appropriate number of free parameters should be selected using an Information Criterion (IC). Using an IC provides an objective and reproducible method for implementing the ``Principle of Parsimony'', i.e. of using the characteristics of the data to find an optimal balance between over- and under-fitting. 

A detailed discussion about IC methods for deciding on the appropriate number of free model parameters for any absorption complex is given in \cite{Webb2021}. That paper compares three ICs: BIC, AICc, and a new information criterion designed specifically for spectroscopy, SpIC. \cite{Webb2021} showed that BIC does not work as well as either AICc or SpIC for absorption system modelling and that SpIC seems to offer significant advantages over AICc. SpIC has been implemented in {\sc ai-vpfit} \citep{Lee2020AI}.

\subsection{Spatial segregation of species} \label{sec:segregation}

The process of tieing redshift parameters between different atomic (or molecular) species raises the question of spatial segregation between different ionisation states or between different elements. The question is of particular importance in the context of measurements of the fine structure constant at high redshift. The Many Multiplet method \citep{Dzuba1999, Webb1999} has been used to search for spectroscopic wavelength shifts between multiple atomic species that may be caused by spacetime variation of the fine structure constant $\alpha$. If different species do not cohabit the same spatial location and if no account is taken of this, measurements of any $\alpha$ variation could be biased. If the same species is used (e.g. Fe\,{\sc ii}), the problem does not exist.

\begin{figure*}
\centering
\includegraphics[width=1.0\linewidth]{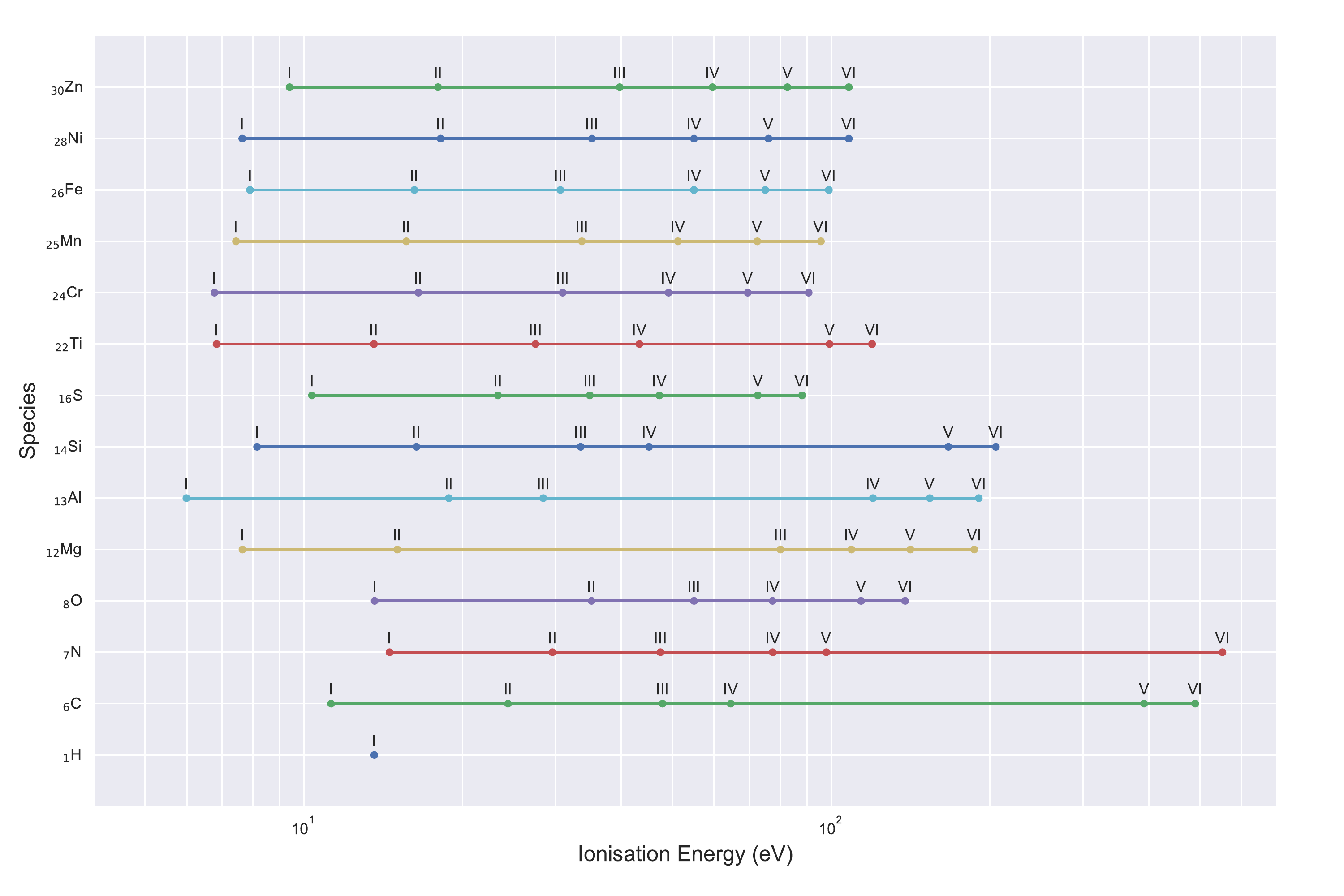}
\caption{Representation of ionisation potentials including absorption species seen in quasar absorption systems. The y-axis shows the element and atomic number. `I' indicates the energy above which one electron is lost, `II' indicates the energy above which two electrons are lost, etc. Figure kindly provided by Vincent Dumont.}
\label{fig:IPs}
\end{figure*}

Fig.~\ref{fig:IPs} provides a graphical representation of species seen in quasar absorption systems, illustrating the range of ionisation potentials. To take one (extreme) example, C\,{\sc i} and C\,{\sc iv} have IPs of 11.3 and 64.5 eV and their relative strengths therefore vary substantially in different physical locations with a galaxy halo, according to local ionisation and other parameters. Should spatial segregation exist between species of widely different atomic number due to gravitational effects, e.g. carbon and zinc, this too could mimic a varying $\alpha$.

{\sc vpfit} correctly accommodates these effects when redshift parameters for the two species are tied because column density parameters are free to iterate. If, for example, there is very little C\,{\sc iv} in a velocity component exhibiting C\,{\sc i} absorption, {\sc vpfit} will reduce the C\,{\sc iv} column density (and discard the line if it falls below the detection threshold). Moreover, any missing components at slightly different redshifts will be discovered by a system like {\sc ai-vpfit} \citep{Lee2020AI}. To summarise the main points of the previous discussion:\\
1. No bias is expected even if all ionisation states are tied at the same redshift; \\ 
2. Free (untied) column density parameters permit species to be discarded if appropriate; \\
3. Where visibly different velocity structures exist e.g. for C\,{\sc iv} and CI, in principle there is no good reason not to tie redshifts in an initial fit because when column densities iterate to a value below the nominal detection thresholds, {\sc vpfit} ``naturally'' allows for this and effectively unties initially tied components. \\
In other words, potential spatial segregation of species is properly accounted for when redshifts are tied, simply because column density parameters are allowed to vary freely.

\subsection{Parameter errors from the Hessian} \label{sec:parerrs}

Informative discussions on parameter error estimation may be found in \cite{Bevington2003, Dahlquist1974, GMW81, Press2007, Irwin1985, Lampton1976}. Once a final model has been found, the unmodified Hessian, Eq.~\eqref{eq:hessianapprox} is used to derive meaningful parameter uncertainties from the Hessian diagonals. The parameter covariance matrix $\hbox{\bf V}$ is then obtained from the inverse of the Hessian matrix \citep{Fisher1958},
\begin{equation}
\mathbfss{V}(\mathbfit{x}) = \mathbfss{G}(\mathbfit{x})^{-1} \, .
\end{equation}

Parameter uncertainties derived from the covariance matrix of course provide no information about additional systematic errors that may be present in real data. If off-diagonal terms in the covariance matrix are zero, the errors in the parameters are independent and $V(\mathbfit{x})_{qq}$ is the estimated variance of the $q$th parameter. Generally however, the off-diagonal terms are non-zero and some level of correlation exists between parameters. The usual approximation is to ignore such correlations, to quote (for simplicity) only diagonal terms as approximate uncertainties, and to check their validity using Monte Carlo calculations and synthetic spectra. This has been done many times for {\sc vpfit} models and in all cases the covariance matrix errors are found to be consistent with (or slight over-estimates compared to) Monte Carlo results, \citet{Webb1987, MurphyPhD2002, King2009, KingPhD2011}.

\subsection{Importance of fitting range selection}

Modelling absorption lines requires a decision as to how much of the flanking continuum regions are included in the fit. The best-fit model parameters are sensitive to this. \cite{Wilczynska2015} carried out a detailed study of 23 absorption systems, comparing the results from different models obtained using varying amounts of continuum flanking regions. The overall results for the sample showed that if the continuum flanking regions were too small, increased scatter was found i.e. additional measurement errors were introduced. 

As far as we know, this effect has not been quantified beyond the \cite{Wilczynska2015} measurements but the implications are obvious: spectral fitting segments should be selected such that they are flanked by line-free regions either side. Damped absorption lines were not looked at in the \cite{Wilczynska2015} study. For non-damped lines, a reasonable rule-of-thumb would be that these regions should be no less than the width of an individual absorption line and preferably larger.

\subsection{Correcting spectral error arrays} \label{sec:correcterrors}

During the numerical procedures used in taking multiple raw astronomical exposures from the telescope to a usable co-added one-dimensional spectrum, re-binning to a common wavelength grid (and other processes) result in small pixel-to-pixel correlations in the final spectrum. It is well known that these effects create a slight smoothing of the data, biasing the value of $\chi^2$ to smaller values. In principle, this could be accommodated by correcting the derivatives of $\chi^2$ with respect to free variables and hence modifying the Hessian matrix and gradient vector accordingly, as discussed in \cite{Irwin1985}. In practice however, this is difficult to implement (because the required noise covariance matrix is generally unknown). The reader is referred to a comprehensive discussion on this problem, and solutions to it, in appendix B ({\it ``Re-binning and combining spectra''}) in the {\sc vpfit} documentation \citep{web:VPFIT}.

\subsection{Linear distortion}

The first check for the presence of potential wavelength distortion effects in echelle quasar spectra was made by \cite{Molaro2008}. Whilst that study did not reveal distortions, improved data showed that for a single exposure, distortion could be identified and that it could be modelled reasonably well using a simple linear function in velocity space \cite{Rahmani2013}. 
Subsequently, \cite{Whitmore2014} applied a simple linear correction to UVES measurements of the fine structure constant by \cite{King2012}. However, a simple linear distortion function is inappropriate in general; the long integration times required to obtain an acceptable signal to noise for quasar spectroscopy require multiple exposures, often taken using different central wavelength settings. This means that in practice, any realistic distortion function is a complex combination of shifted linear functions and rarely resembles a simple linear relationship. For this reason, a detailed analysis was reported in \cite{Dumont2017}, describing more appropriate distortion functions, determined on a case by case basis.

The \cite{Dumont2017} distortion modelling was carried out as an external analysis, separate from the {\sc vpfit} process. The main disadvantage of externally solving for distortion was one of calculation time; each quasar absorption system required a large set of models, minimising the overall system $\chi^2$ as a function of the distortion model parameters. The \cite{Dumont2017} model has now been implemented within {\sc vpfit} such that all parameters, i.e. absorption parameters as well as distortion parameters, are solved for simultaneously. The approximation made is that the velocity shift pattern can be described by a single parameter $\gamma$, the slope of the linear distortion model,
\begin{equation}
    v_{dist,i}(\lambda) = \gamma (\lambda - \lambda_{cent,i}) \, ,
\end{equation}
where $\lambda_{cent,i}$ is the central wavelength of the $i^{th}$ exposure. Whilst this approximation is somewhat unsatisfactory, we have little choice because a model adopting one slope for each exposure used in the co-added spectrum would create modelling degeneracies. The overall velocity shift is the weighted linear combinations of $v_{dist,i}(\lambda)$,
\begin{equation}
    v_{net} = \sum_i \frac{\sqrt{T_i} v_{dist,i}(\lambda)}{\sqrt{T_i}} \, ,
\end{equation}
where $\sqrt{T_i}$ is the square root of exposure time of the observational segment, thus playing the role of the weight for the $i^{th}$ exposure. The distortion slope parameter $\gamma$ is taken as a free fitting parameter and its statistical uncertainty is derived from the Hessian matrix at the best fit in {\sc vpfit}.

\bsp
\label{lastpage}
\end{document}